\documentclass[prr,superscriptaddress,twocolumn]{revtex4-1}
\usepackage[latin9]{inputenc}
\usepackage{color}
\usepackage{amsmath}
\usepackage{amsfonts}
\usepackage{amssymb}
\usepackage{graphicx}
\usepackage[unicode=true,
 bookmarks=true,bookmarksnumbered=false,bookmarksopen=false,
 breaklinks=false,pdfborder={0 0 1},backref=false,colorlinks=true]
 {hyperref}
\hypersetup{
 linkcolor=magenta,urlcolor=blue,citecolor=blue,pdfstartview={FitH},urlcolor=blue}

\makeatletter
\usepackage{amsmath}
\usepackage{amssymb}
\usepackage{graphicx}
\usepackage{graphbox}

\usepackage{amsfonts}
\usepackage{tabularx}
\usepackage{dcolumn}
\usepackage{bm}
\usepackage{epstopdf}
\usepackage{times}

\newcommand{\ket}[1]{|#1\rangle}
\newcommand{\bra}[1]{\langle #1 |}

\newcommand{\pardif}[2]{\frac{\partial #1}{\partial #2}}

\makeatother

\begin{document}
\title{Quantum Information Scrambling in Quantum Many-body Scarred Systems}
\author{Dong Yuan}
\thanks{These authors contributed equally to this work.}
\affiliation{Center for Quantum Information, IIIS, Tsinghua University, Beijing 100084, People's Republic of China}
\author{Shun-Yao Zhang}
\thanks{These authors contributed equally to this work.}
\affiliation{Center for Quantum Information, IIIS, Tsinghua University, Beijing 100084, People's Republic of China}

\author{Yu Wang}
\affiliation{Department of Physics, Harvard University, Cambridge, Massachusetts 02138, USA}

\author{L.-M. Duan}
\email{lmduan@tsinghua.edu.cn}
\affiliation{Center for Quantum Information, IIIS, Tsinghua University, Beijing 100084, People's Republic of China}

\author{Dong-Ling Deng}
\email{dldeng@tsinghua.edu.cn}
\affiliation{Center for Quantum Information, IIIS, Tsinghua University, Beijing 100084, People's Republic of China}
\affiliation{Shanghai Qi Zhi Institute, 41st Floor, AI Tower, No. 701 Yunjin Road, Xuhui District, Shanghai 200232, China}

\begin{abstract}
Quantum many-body scarred systems host special non-thermal eigenstates that support periodic revival dynamics and weakly break the ergodicity. Here, we study the quantum information scrambling dynamics in quantum many-body scarred systems, with a focus on the ``PXP" model. We use the out-of-time-ordered correlator (OTOC) and Holevo information as measures of the information scrambling, and apply an efficient numerical method based on matrix product operators to compute them up to $41$ spins. We find that both the OTOC and Holevo information exhibit a linear light cone and periodic oscillations inside the light cone for initial states within the scarred subspace, which is in sharp contrast to thermal or many-body localized systems. The periodic revivals of OTOCs and Holevo information signify unusual breakdown of quantum chaos and are not equivalent to the revival dynamics of state fidelity or local observables studied in the previous literature.
To explain the formation of the linear light cone structure, we provide a perturbation-type calculation based on a phenomenological model. In addition, we demonstrate that the OTOC and Holevo information dynamics of the ``PXP" model can be measured using the Rydberg-atom quantum simulators with current experimental technologies, and numerically identify the measurable signatures using experimental parameters.
\end{abstract}

\maketitle
\section{Introduction}
\begin{figure}
\hspace*{-0.48\textwidth}
\includegraphics[width=0.48\textwidth]{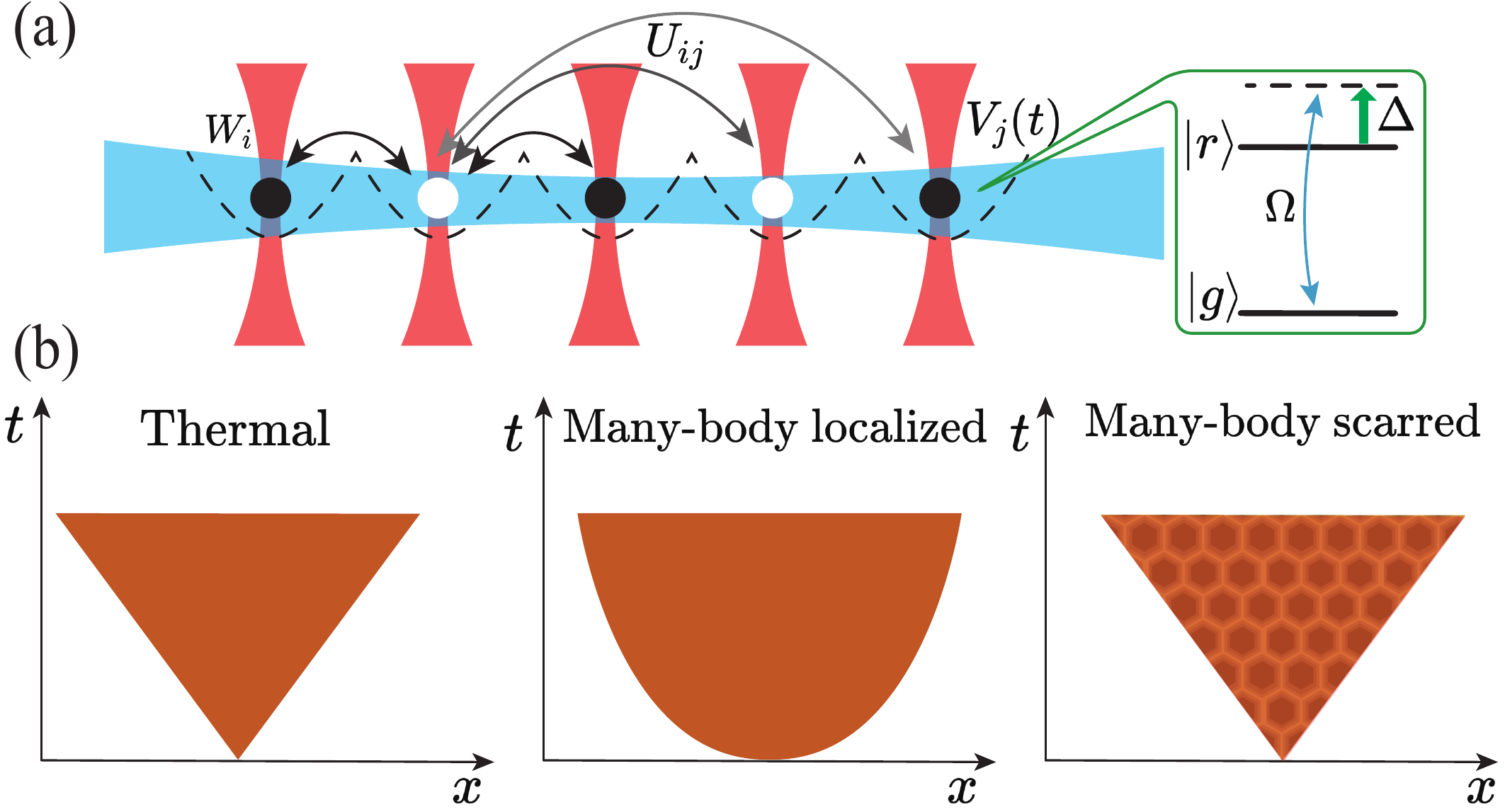} 
\caption{
(a) An illustration of the one dimensional Rydberg atom array used for the measurement of OTOCs $F_{ij}(t)$ and Holevo information $\chi_j(t)$. Individual neutral atoms are trapped with optical tweezers (vertical red beams). The global Rydberg laser (the horizontal blue beam), with the Rabi frequency $\Omega$ and detuning $\Delta$, couples the atomic ground state $\ket{g}$ (black circles) with the Rydberg excited state $\ket{r}$ (white circles). A pair of Rydberg atoms at the distance $R_{ij}$ shares the van der Waals repulsion $U_{ij}=U_0/R_{ij}^6$. 
(b) A schematic illustration of the information scrambling dynamics in systems of different thermalization classes. 
}
\label{fig:illustration} 
\end{figure}

Isolated quantum many-body systems would eventually thermalize under the time evolution and their subsystems relax to the equilibrium, leading to the emergence of ergodicity and statistical mechanics. During this process, any local information preserved in the initial states scrambles into the entire system and becomes unrecoverable. This kind of quantum thermalization phenomena has been illustrated by the eigenstate thermalization hypothesis (ETH) in the past decades \cite{Deutsch1991Quantum,Srednicki1994Chaos}. While numerous works have confirmed the universality and correctness of ETH in various scenarios \cite{Deutsch2018Eigenstate,Kim2014testing,Deutsch2013Microscopic,Rigol2008Thermalization}, discovering quantum many-body systems violating the ETH is still of fundamental and practical importance. Known exceptions to the ETH paradigm include the integrable \cite{Sutherland2004beautiful} and many-body localized (MBL) \cite{Nandkishore2015Manybody,Abanin2019Colloquium} systems, which have either exact or approximate extensive conserved quantities to prevent the systems from thermalization. Recently, in experiments with Rydberg atoms, non-thermal periodic revival dynamics has been observed after quenching the system from a special high-energy N\'eel state \cite{Bernien2017Probing,Bluvstein2021Controlling}. Subsequent theoretical works attribute this weak ergodicity breaking to a small fraction of ETH-violating eigenstates (dubbed ``quantum many-body scars")  embedded in a sea of thermal eigenstates \cite{Turner2018weak,Turner2018quantum,Serbyn2021quantum} . Here, we investigate the scrambling of quantum information in many-body scarred systems (see Fig. \ref{fig:illustration} for a pictorial illustration).

Quantum information scrambling describes the propagation and effective loss of initial local information in quantum many-body dynamics. It has  attracted considerable attention in different contexts, including the black hole thermaldynamics \cite{Hayden2007Black,Shenker2014black,Maldacena2016bound}, quantum many-body quench dynamics \cite{Lewis2019dynamics,Swingle2018unscrambling,Richerme2014Nonlocal,Jurcevic2014Quasiparticle}, and machine learning \cite{Shen2020Information,Wu2021Scrambling,Garcia2021Quantifying}. Except for a few pronounced integrable examples \cite{Lin2018Out,Gopalakrishnan2018Operator,Khemani2018Velocity,McGinley2019Slow,Fortes2019Gauging,xu2020accessing,Xu2020Does,Goldfriend2020Quasi,Riddell2021scaling,Lopez2021Operator,Shukla2022Out}, for general quantum many-body systems with short-range interactions obeying the ETH, the Lieb-Robinson bound \cite{Lieb1972Finite} restricts the correlation propagation within a linear light cone (analogous to the causal light cone in relativistic theories). In contrast, strong disorders in MBL systems prevent the local transport, resulting in a logarithmic light cone for information spreading \cite{Deng2017Logarithmic,Huang2017out,Fan2017out,Chen2016universal,Chen2017out,Banuls2017Dynamics} [see Fig. \ref{fig:illustration}(b) for a sketch]. Quantum many-body scars, as a new thermalization class in many-body dynamics, possess the potential to exhibit different information spreading behaviours from the former two cases. Despite previous extensive studies of quantum many-body scars from various perspectives \cite{Choi2019emergent,Ho2019periodic,Michailidis2020Slow,Turner2021Correspondence,Chattopadhyay2020quantum,Maskara2021discrete,Mukherjee2020Collapse,Moudgalya2018Exact,Moudgalya2018Entanglement,Schecter2019Weak,Khemani2019Signatures,Iadecola2019Quantum,Moudgalya2020eta,Mark2020eta,Desaules2021Proposal,Scherg2021Observing,Langlett2021Hilbert,Moudgalya2021Quantum,Lee2020Exact,Jeyaretnam2021Quantum,Lin2020Slow,Surace2021Exact,Shem2021Fate,Huang2021Stability,Surace2021Quantum,Ren2021Quasisymmetry,Robust2021Dooley,langlett2021rainbow,yao2021quantum}, the exploration of quantum information scrambling in quantum many-body scarred systems is still lacking hitherto.

In this paper, we apply the out-of-time-ordered correlator (OTOC) and Holevo information as measures to study this problem. 
We find that both the OTOC and Holevo information exhibit a linear light cone and periodic revival dynamics inside the light cone for initial states within the scarred subspace, displaying distinct features from the ETH and MBL cases [Fig. \ref{fig:illustration}(b)]. Moreover, the persistent oscillations will disappear and the information propagation speed will increase once we choose generic high-energy initial states. 
The periodic oscillations of OTOCs and Holevo information signify unusual breakdown of quantum chaos and persistent backflow of quantum information.
Due to the action of interleaved operators, the dynamics of OTOCs and Holevo information actually involve both the scarred subspace and the thermal eigenstate bath. We emphasize that their dynamics can not be readily deduced from the eigenstate decomposition of initial states, thus distinguishing our work from the previous literature \cite{Turner2018weak,Turner2018quantum,Serbyn2021quantum}.
In order to explain the linear light cone structure, we further provide a perturbation-type calculation in the interaction picture based on a phenomenological model proposed in \cite{Choi2019emergent}. Finally, we propose an experiment with Rydberg atoms to observe the predicted exotic OTOC and Holevo information dynamics, and numerically identify the measurable signatures using experimental parameters.

\section{Model} 
Motivated by the Rydberg-atom experiment \cite{Bernien2017Probing}, in the limit of Rydberg blockade \cite{Saffman2010Quantum} the physics of quantum many-body scars are extracted as the one-dimensional (1D) ``PXP" model with an open boundary condition \cite{Turner2018weak,Turner2018quantum}:  
\begin{equation}
    H = \sum_{j=1}^L P_j \sigma^x_{j+1} P_{j+2},
\label{Eq:PXP}
\end{equation}
where $P_j = (1-\sigma^z_j)/2$, $\sigma^{x,y,z}_j$ are Pauli matrices of the $j$-th qubit, $L$ is the number of total qubits, and $\ket{\downarrow(\uparrow)}$  represents the atomic ground (Rydberg excited) state $\ket{g(r)}$. 
Below we always consider the dynamics within the constrained Hilbert space (where computational bases with two nearby up spins $\ket{\cdots \uparrow\uparrow \cdots }$ are removed). 
The PXP Hamiltonian is non-integrable and chaotic according to the level statistics studies, yet it holds a small fraction of ETH-violating scarred eigenstates that support the periodic revival dynamics for initial states within the scarred subspace (such as the  N\'eel state $\ket{Z_2}=\ket{\uparrow\downarrow\uparrow\cdots\uparrow\downarrow}$). Whereas, for generic high-energy initial states (such as $\ket{\bm{0}}=|\downarrow\downarrow\cdots\downarrow\rangle$) the dynamics will quickly become chaotic and no revival occurs \cite{Turner2018weak,Turner2018quantum,Ho2019periodic}.

The OTOC utilizes the Heisenberg operator growth to characterize the information scrambling and quantum chaos, and is defined as \cite{Larkin1969Quasiclassical,Maldacena2016bound,Hashimoto2017out}
\begin{equation}
    F_{ij}(t) = \langle \psi | W_i^\dagger V_j^\dagger(t) W_i V_j(t) |\psi\rangle,
\label{Eq:OTOC}
\end{equation}
where $|\psi \rangle $ is an initial pure state, $W_i, V_j$ are local observables defined on sites $i,j$, and $V_j(t)=e^{iHt} V_j e^{-iHt}$ ($\hbar=1$). The OTOC directly connects to the squared commutator $C_{ij}(t) = \langle \psi | [W_i,V_j(t)]^\dagger [W_i,V_j(t)] |\psi\rangle$ by the relation $C_{ij}(t) = 2(1 - {\rm{Re}} (F_{ij}(t)))$, for unitary operators $W_i,V_j$. A simple physical picture for the OTOC is that if the Heisenberg operator growth of $V_j(t)$ does not reach the site $i$, $[W_i,V_j(t)]=0, F_{ij}(t)=1$, while the equalities will break down when sites $i,j$ become correlated inside the causal region. 
Note that the OTOCs $F_{ij}(t)$ consist of the forward and backward Hamiltonian evolution, between which there exist interleaved local operators $W_i$ and $V_j$. Since the evolved quantum states ($e^{-i H t} \ket{\psi}$ or $e^{-i H t} W_i \ket{\psi}$) in general are not the eigenstates of inserting operators $W_i$ and $V_j$, the OTOC dynamics can be essentially different from those of simple local observables. For instance, in the study of MBL systems, people have realized the distinction between the dynamics of OTOCs and common two-point correlators: OTOCs exhibit a logarithmic light cone and decay inside \cite{Deng2017Logarithmic,Huang2017out,Fan2017out,Chen2016universal,Chen2017out,Banuls2017Dynamics}, whereas the two-point correlators are bounded to be exponentially small due to the localization nature \cite{Nandkishore2015Manybody,Abanin2019Colloquium}. In the following discussions, we mainly focus on the $ZZ$-OTOC ($W=\sigma^z, V=\sigma^z$) and $XZ$-OTOC ($W=\sigma^x, V=\sigma^z$).

The Holevo information (or Holevo $\chi$ quantity) originates from the quantum information theory to upper bound the accessible information between two separate agents \cite{Holevo1973bounds,Nielsen2010Quantum}. Consider that Alice prepares mixed states $\rho_X$ in the set $\{\rho_1,\rho_2,\cdots,\rho_n\}$ with probability $\{p_1,p_2,\cdots,p_n\}$ respectively, and then sends $\rho_X$ to Bob. With any kind of positive operator-valued measures (POVMs), the amount of information Bob can obtain about the variable $X$ according to the measurement outcome $Y$ is bounded by $I(X:Y) \le \chi = S(\sum_{i=1}^n p_i \rho_i)-\sum_{i=1}^n p_i S(\rho_i)$, where $I(X:Y)$ is the mutual information between $X$ and $Y$, $S(\rho)=-\textnormal{Tr}(\rho \log \rho)$ denotes the von Neumann entanglement entropy. Roughly speaking, the Holevo information describes the distinguishability of states in the set $\{\rho_1,\rho_2,\cdots,\rho_n\}$. For instance, $p_1=p_2=1/2$, if $\rho_1=\ket{\uparrow}\bra{\uparrow},\rho_2=\ket{\downarrow}\bra{\downarrow}$, then $\chi=1$; while if $\rho_1=\rho_2=\ket{\uparrow}\bra{\uparrow}$, then $\chi=0$. Compared with OTOCs, Holevo information possesses richer information-theoretic meanings and its experimental measurement does not require the inverse Hamiltonian evolution. We hence expect that Holevo information can be used to characterize the information spreading dynamics and solve information-theoretic problems in more quantum many-body systems \cite{Bao2017Distinguishability,Guo2018Distinguishing,Qi2022Holevo}.

Here, we regard the reduced Hamiltonian evolution on subsystems as quantum communication channels in the original setup of Holevo information \cite{Holevo1973bounds} and use it to study the information scrambling dynamics.
Consider the Hamiltonian evolution on two different initial states, $\ket{\psi}$ and $\sigma^x_i \ket{\psi}$, taking $\ket{\psi}$ to be a computational basis state (e.g., $\ket{Z_2}$ or $\ket{\bm{0}}$). In other words, initially we have encoded one bit  information at the site $i$ [$\ket{\uparrow(\downarrow)}_i$]. 
We demonstrate how the one-bit information scrambles into the entire system by computing the Holevo information on the site $j$:
\begin{equation}
    \chi_j(t) = S\left(  \frac{\rho_j(t) +\rho'_j(t)}{2} \right) -  \frac{S\left(\rho_j(t) \right) + S\left(\rho'_j(t) \right)}{2},
\label{Eq:Holevo}
\end{equation}
where $\rho_j(t)$ and $\rho^\prime_j(t)$ are reduced density matrices of the $j$-th spin after the Hamiltonian evolution for the initial state $\ket{\psi}$ and $\sigma^x_i \ket{\psi}$ respectively.
$\chi_j(t)$ measures how much information one could obtain by any local probe on the $j$-th site for these two sets of evolution.

\begin{figure*}
\hspace*{-0.94\textwidth}
\includegraphics[width=0.94\textwidth]{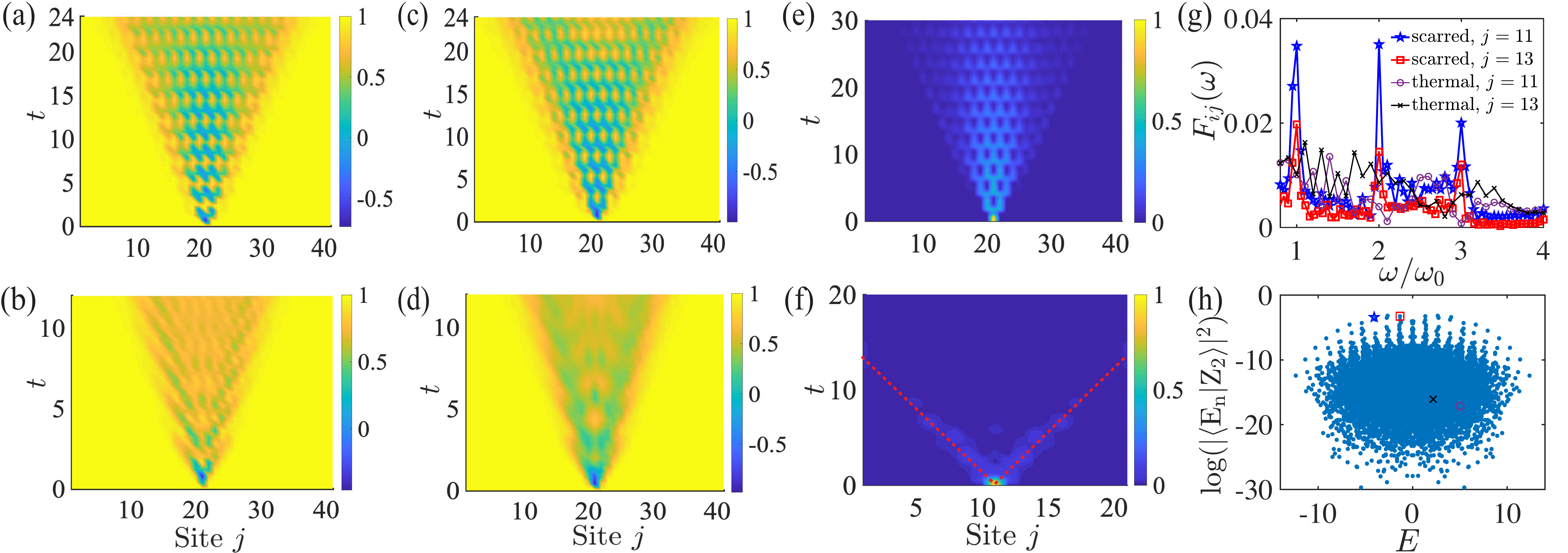} 
\caption{Information scrambling dynamics in quantum many-body scarred systems. Spatio-temporal evolution of the $ZZ$-OTOCs (a), (b) and $XZ$-OTOCs (c), (d) of the PXP model, calculated by the MPO method, with $L=41$, and $i=21$. The OTOCs $F_{ij}(t)$ exhibit linear light cones and periodic oscillations in (a), (c) with the initial state $|Z_2\rangle$, in contrast to the larger butterfly velocity and absence of oscillations in (b), (d) with the initial state $\ket{\bm{0}}$. (e), (f) The spatio-temporal evolution of the Holevo information $\chi_j(t)$ for the initial state $|Z_2\rangle$ ($L=41$, by MPS) and  $\ket{\bm{0}}$ [$L=21$, by exact diagonalization (ED)]. Initially, one-bit information is encoded in the central qubit. The information travels ballistically and oscillates in (e), while peaks and quickly diminishes in (f) (the peaks of $\chi_j(t)$ for $j$ far from the central qubit are too small to be distinguished; the red dashed lines are a guide to eye). 
All the color maps are interpolated to non-integer $j$ to better illustrate the light cone. (g) The frequency spectra of the $ZZ$-OTOC dynamics for the scarred eigenstates (marked by star and square) and thermal eigenstates (marked by circle and cross). (h) The log-scaled overlap between each eigenstate $\ket{E_n}$ and the $\ket{Z_2}$ N\'eel state, $L=20$, calculated by ED. 
}
\label{fig:numerics} 
\end{figure*} 

\section{Numerical simulations}
In Fig. \ref{fig:numerics}, we numerically compute the OTOC and Holevo information for the PXP model as diagnoses of the information scrambling dynamics. Specifically, we apply an efficient matrix-product-operator (MPO) method \cite{xu2020accessing} to calculate the $ZZ$-OTOC and $XZ$-OTOC up to $L=41$ spins (see Appendix. \ref{sec:more numerics} for algorithm details). We observe the following features: For the initial state $\ket{\psi}=\ket{Z_2}$ [Fig. \ref{fig:numerics}(a), (c)], both the $ZZ$-OTOC and $XZ$-OTOC spread ballistically, forming a linear light-cone structure with the butterfly velocity $v_b\sim 0.6$ (inverse of the light cone slope) \cite{Shenker2014black,Roberts2016Lieb,stahl2018asymmetric,zhang2020asymmetric,Liu2018Asymmetric,Zhang2021Anomalous}; inside the light cone, the OTOC dynamics show evident periodic revivals with the period $T\approx 4.71$, consistent with the oscillation period of state fidelity and local observables in Hamiltonian evolution \cite{Turner2018weak,Ho2019periodic}. Besides, the oscillations for different sites $j$ are \textit{synchronized}, meaning that $F_{ij}(t)$ of different $j$ have the same period $T$ and reach the maxima at the same time $t$. In contrast, for a generic high-energy initial state like $\ket{\bm{0}}$ (Fig. \ref{fig:numerics}(b), (d)), the OTOCs have a larger butterfly velocity $v_b\sim 1$ (information scrambles faster) and decay without discernible periodic revivals inside the light cone.

We emphasize that except for the $ZZ$-OTOC of the $\ket{Z_2}$ initial state, the periodic oscillations of OTOCs inside the light cone can not be deduced from the approximate recurrence of the $\ket{Z_2}$ state under Hamiltonian evolution. For instance, in the $XZ$-OTOC case, the inserting $\sigma_i^x$ operators will flip the $\ket{Z_2}$ state partially out of the scarred subspace and introduce the influence from the thermal eigenstate bath. In order to confirm the generality of the revival behaviours, we numerically show that the persistent and synchronized oscillations still appear even if the initial states are replaced with scarred \textit{eigenstates}. We use the overlap between eigenstates $\ket{E_n}$ of the PXP Hamiltonian and $\ket{Z_2}$ [Fig. \ref{fig:numerics}(h)], as well as the half-chain entanglement entropy of $\ket{E_n}$ (see Appendix. \ref{sec:more numerics} for details), to distinguish quantum many-body scars from typical thermal eigenstates \cite{Turner2018weak,Turner2018quantum}. In Fig. \ref{fig:numerics}(g), we transform the $ZZ$-OTOC dynamics into the frequency domain and find that: For initial states being scarred eigenstates, there exist $\omega_0=2\pi/T, 2\omega_0, 3\omega_0$ peaks in the spectra for different sites $j$, but not for the case of initial states being the generic thermal eigenstates. Fig. \ref{fig:numerics}(g) indicates that the periodic revivals of OTOCs are general phenomena for initial states within the scarred subspace. The recurrence of quantum information signified by the OTOCs is not equivalent to the recurrence of quantum states through Hamiltonian evolution, for which the energy eigenstates actually have \textit{no} dynamics. In Appendix. \ref{sec:more numerics} we further display the OTOC dynamics with initial states being eigenstates of the PXP Hamiltonian and superposition states of $\ket{Z_2}$ and $\ket{Z_2'}=(\prod_{i=1}^L \sigma_i^x) \ket{Z_2}$ .

Similar information scrambling dynamics also emerge when probed by the Holevo information. As shown in Fig. \ref{fig:numerics}(e), for two sets of Hamiltonian evolution on $\ket{Z_2}$ and $\sigma_{\lceil L/2 \rceil}^x \ket{Z_2}$, $\chi_j(t)$ initially vanishes everywhere except on the central qubit, where one bit local information is encoded. As time evolves, non-zero Holevo information can be probed at other sites $j$, which forms a linear light cone in spacetime and also persistent oscillations inside the light cone with a period $T\approx 4.73$, consistent with the OTOC results. In comparison with the initial states $\ket{\bm{0}}$ and $\sigma_{\lceil L/2 \rceil}^x\ket{\bm{0}}$ [Fig. \ref{fig:numerics}(f)], $\chi_j(t)$  peaks when the information wavefront arrives and eventually diminishes for long time $t$, following the indistinguishability of $\rho_j(t)$ and $\rho^\prime_j(t)$ predicted by the ETH. 
The information spreading velocity $v_h$ for the $\ket{Z_2}$ case is $v_h\sim 0.6$, less than that of the $\ket{\bm{0}}$ case $v_h\sim 0.9$. 
Similarly, due to the action of the $\sigma_{\lceil L/2 \rceil}^x$ operator, the Holevo information dynamics involve both the scarred subspace and the thermal eigenstate bath, which are further discussed in Appendix. \ref{sec:Holevo}.
Fig. \ref{fig:numerics}(e) is obtained by the time-evolving block decimation (TEBD) algorithm \cite{vidal2003efficient,schollwock2011density}  based on the matrix-product-state (MPS) ansatz, which leverages the relatively low entanglement entropy of scarred eigenstates and is not applicable to the $\ket{\bm{0}}$ case \cite{Turner2018weak,Turner2018quantum}.

\section{Analytical explanations}
In this section, we provide analytical explanations for the OTOC and Holevo information dynamics observed in numerical simulations. First, we mention that the analytical computation of OTOCs is a challenging task for general quantum many-body systems, despite a few pronounced solvable examples such as the Sachdev-Ye-Kitaev (SYK) model \cite{Sachdev1993Gapless,kitaev2015simple,Gu2017local,Gu2021Twoway}. For the PXP model, it is difficult to directly deal with its OTOC dynamics by the perturbation method since all the terms in the Hamiltonian have the same interaction strength. However, the essential features of the PXP model consist of two parts: the periodic oscillations within and the chaotic dynamics out of the scarred subspace, which we adopt a phenomenological model proposed in \cite{Choi2019emergent} to effectively describe: $H' =(\Omega/2) \sum_{i} \sigma_{i}^{x}+\sum_{i} R_{i, i+3} P_{i+1, i+2}$, where $P_{i,i+1}=(1-\vec{\sigma}_i\cdot \vec{\sigma}_{i+1})/4$, $R_{i,j}=\sum_{\mu,\nu}J^{\mu\nu}_{ij}\sigma^\mu_i \sigma^\nu_j$ ($J^{\mu\nu}_{ij}$ are random coupling constants, $\mu,\nu=\{x,y,z\}$) for a 1D chain with $L$ spins. 
$H_0'=(\Omega/2) \sum_{i} \sigma_{i}^{x}$ corresponds to the periodic rotations in the PXP Hamiltonian while $R=\sum_{i} R_{i, i+3} P_{i+1, i+2}$ plays the role of thermalization for states out of the scarred subspace.
The $L+1$ scarred eigenstates are all the $x$-direction Dicke states $H' \ket{s=L/2, S^x=m_x} = m_x \Omega \ket{s=L/2, S^x=m_x}, \ (m_x=-s,-s+1,\cdots,s-1,s)$. Once we start the Hamiltonian evolution from $\ket{\psi}=\ket{\uparrow\uparrow\cdots\uparrow}$, perfect quantum state revivals with period $T=2\pi/\Omega$ will be observed, imitating most but not all the characteristics of the PXP model \cite{Choi2019emergent} (see Appendix. \ref{sec:analytical} for model details and relevant numerical results).

For the non-trivial situation of $XZ$-OTOC $W_1=\sigma^x_1$, $V_r=\sigma^z_r$ with the initial state $\ket{\psi}=\ket{\uparrow\uparrow\cdots\uparrow}$, specially at time points $t=n T/2\ (n=0,1,2\cdots)$, the OTOCs are simplified into $F(r,t)=(-1)^n\bra{\phi(t)} \sigma^z_r \ket{\phi(t)}$, where $\ket{\phi(t)}=e^{-iH' t} \sigma^x_1\ket{\psi}$. The physical picture for the formation of linear light cone structure in OTOC and Holevo information dynamics is that: Without the action of $\sigma^x_1$, the state $\ket{\phi(t)}$ will undergo perfect periodic oscillations, thus $F(r,t=n T/2)\equiv 1$. However, now $\sigma^x_1$ penetrates the scarred subspace on the first site. The ``heat flow" (like a quasi-particle created by the local quench \cite{Jurcevic2014Quasiparticle}) leaks  and propagates ballistically through the entire system, leading to the decay of OTOC $F(r, t = n T/2) < 1$ after the wavefront of heat flow arrives. Note that in the setup of Holevo information dynamics, we also have a Hamiltonian evolution term $e^{-i H t} \sigma_{\lceil L/2 \rceil}^x \ket{Z_2}$, which has the same structure of $\ket{\phi(t)}$. We hence deduce that the dynamics of Holevo information and OTOCs are closely related and follow the same physical picture. Indeed the numerical results of the two criteria support each other and exhibit similar information scrambling behaviors.

We use the interaction picture of $H_0'$ to remove the Rabi oscillation effect. For the early growth region of OTOCs, we can split the evolution time $t$ into $r$ pieces with $\Delta t=t/r,\  J\Delta t \ll 1$, where $J$ is the average energy scale of all the $J^{\mu\nu}_{ij}$. After some lengthy calculations shown in Appendix. \ref{sec:analytical}, the deterioration of perfect oscillations at the site $r$ will be dominantly caused by an operator product series $\hat{R}^{r}\left( (r-1)\Delta t\right) \cdots \hat{R}^2(\Delta t) \hat{R}^1(0)$, [$\hat{R}^i (n \Delta t)= e^{i H_0' n\Delta t} R_{i, i+3} P_{i+1, i+2} e^{-i H_0' n\Delta t}$ are operators in the $H_0'$ interaction picture], leading to 
\begin{equation}
    F(r,t)\sim 1 - (\frac{a J t}{r})^r.
\label{Eq:analytical}
\end{equation}
Here, $a$ is some model-dependent $O(1)$ constant and we have generalized the special time points $t=n T/2$ to arbitrary $t$ before the wavefront arrives. Eq. \eqref{Eq:analytical} depicts a linear light cone $t \propto r/J$ in the early growth region of OTOCs.

Despite the fact that the OTOC dynamics have linear light cone structure both in the ETH and many-body scarred systems, the thermalization processes are pretty different: In the ETH systems, when we start the Hamiltonian evolution, thermalization happens everywhere \textit{globally} on the quantum state; On the contrary for many-body scarred systems, according to the calculations above, the periodic revivals are destroyed \textit{locally} at some sites and the deterioration effect then propagates through the entire system. The global versus local thermalization processes are reminiscent of the spin correlation dynamics in global \cite{Richerme2014Nonlocal} versus local \cite{Jurcevic2014Quasiparticle} quench dynamics, and can also explain the larger butterfly velocity for the $\ket{\bm{0}}$ initial state than that of the $\ket{Z_2}$ initial state in OTOC and Holevo information dynamics. 

The perturbation-type calculations break down for the OTOC and Holevo information dynamics deep inside the light cone. As mentioned in the previous section, the persistent and synchronized oscillations inside the light cone still appear even if the initial states are replaced with scarred eigenstates (see Appendix. \ref{sec:more numerics} Fig. \ref{fig:otherscarZZ} and \ref{fig:otherscarZZclose}). We attribute the unusual revival dynamics inside the light cone to the local rather than global thermalization argued above and robustness of scarred eigenstates against local perturbations: The quantum many-body scars have been shown to present certain robustness against perturbations and disorders \cite{Lin2020Slow,Surace2021Exact,Shem2021Fate,Huang2021Stability,Surace2021Quantum}. Under some local perturbations like $\sigma^{x(z)}_{i(j)}$ operators on the evolved quantum states ($e^{-i H' t}\sigma^x_1 \ket{\psi}$), thermalization happens locally and propagates outwards. After the wavefront of thermalization effect passes, the periodic oscillations are preserved to some extent due to the robustness of scar eigenstates, leading to the revival pattern observed in the OTOC and Holevo information dynamics. This picture are also observed in Appendix. \ref{sec:analytical} Fig. \ref{fig:OTOC_toy_SM} and Appendix. \ref{sec:Holevo} Fig. \ref{fig:vec_rot}, and could be an interesting avenue for future investigations.

\section{Experiment proposal} 
Recent experimental progress has enabled the measurements of information scrambling dynamics in well-controlled synthetic quantum systems, including nuclear magnetic resonance systems \cite{Li2017Measuring,Nie2020Experimental}, trapped ions \cite{garttner2017measuring,landsman2019verified,Joshi2020Quantum,Green2021Experimental} and superconducting qubits \cite{mi2021information,Blok2021Quantum,braumuller2021probing,zhu2021observation}. The PXP model can be naturally realized with the Rydberg-atom platform \cite{Saffman2010Quantum,Saffman2016Quantum,browaeys2020many}, governed by the following Hamiltonian:
\begin{equation}
    \mathcal{H}=\frac{\Omega}{2}\sum_{i} \sigma^{x}_{i} - \Delta\sum_{i} n_{i} + \sum_{i < j} U_{ij} n_{i} n_{j}.
\label{Eq:RydbergHam}
\end{equation}
$\sigma_i^x$ connects the atomic ground ($\ket{g}$) and Rydberg excited ($\ket{r}$) state on the site $i$ with the Rabi frequency $\Omega$ and detuning $\Delta$. Two Rydberg atoms at distance $R_{ij}$ share the van der Waals repulsion $U_{ij} = U_0/R_{ij}^6$. $n_i=(1+\sigma^z_i)/2$ denotes the number of Rydberg states. In the parameter regime of Rydberg blockade $U_{i,i+1}=U_1 \gg \Omega \gg U_{i,i+2}=U_2,\ \forall i$, the lowest $U(1)$ symmetry sector $\sum_i n_i n_{i+1} = 0$ of Hamiltonian Eq. \eqref{Eq:RydbergHam} approximately reduces to the PXP model, which provides us an opportunity to experimentally measure the $ZZ$-OTOC and Holevo information dynamics [see Fig. \ref{fig:illustration}(a)].

The main difficulty in simulating OTOC dynamics lies in the implementation of the inverse Hamiltonian evolution $\exp{\left(-i(-H)t\right)}$. Fortunately, the PXP model has a particle-hole symmetry operation $ \left(\prod_i \sigma_i^z\right) H \left(\prod_i \sigma_i^z\right) = -H$ to reverse the sign of $H$. For the Rydberg-atom Hamiltonian Eq. \eqref{Eq:RydbergHam}, the $\prod_i \sigma_i^z$ operator only changes the sign of the Rabi oscillation term while keeping the detuning term and Rydberg blockade structure intact. Below we denote the Hamiltonians for experimental evolution and inverse evolution as $\mathcal{H}_{\pm} = \pm (\Omega/2) \sum_{i}  \sigma^{x}_{i} - \Delta \sum_{i} n_{i} + U_1 \sum_i  n_{i} n_{i+1} + U_2 \sum_i  n_{i} n_{i+2}$, where we only take the next-nearest-neighbor interaction $U_2$ into consideration of error analysis due to the sixth power decay. Specifically, for the case of $ZZ$-OTOC $F_{ij}(t) = \langle \psi | \sigma_i^z \sigma_j^z(t) \sigma_i^z \sigma_j^z(t) |\psi\rangle$ and initial states $\ket{\psi}=\ket{Z_2}$ or $\ket{\bm{0}}$, since $\sigma_i^z \ket{\psi}= (-1)^{n_i+1} \ket{\psi}$, we have
\begin{equation}
    F_{ij}(t) = (-1)^{\bra{\psi} n_i \ket{\psi}+1} \langle \varPsi_j(t) | \sigma^z_i | \varPsi_j(t) \rangle,
\label{Eq:ob_ave_exp}
\end{equation}
where $| \varPsi_j(t) \rangle = e^{-i\mathcal{H}_{-} t} \sigma^z_j e^{-i\mathcal{H}_{+} t} |\psi \rangle$, and $e^{-i\mathcal{H}_{-} t} = \left(\prod_i \sigma_i^z\right) e^{-i\mathcal{H}_{+} t} \left(\prod_i \sigma_i^z\right)$. The measurements of $ZZ$-OTOC are hence reduced to the evaluation for the expectation value of $\sigma_i^z$  on a time-dependent quantum state $\ket{\varPsi_j(t)}$. The implementation of $\ket{\varPsi_j(t)}$ only requires Hamiltonian evolution of Eq. \eqref{Eq:RydbergHam} together with global and individual Pauli-$Z$ gates \cite{Saffman2010Quantum,Saffman2016Quantum}.
The measurement protocol for the Holevo information is straightforward: preparing two initial states $\ket{Z_2(\bm{0})}$ and $\sigma_{\lceil L/2 \rceil}^x\ket{Z_2(\bm{0})}$, evolving the system with $\mathcal{H}$ independently, and finally doing quantum state tomography for all the single-qubit density matrices to obtain $\chi_j(t)$ via Eq. \eqref{Eq:Holevo}.

\begin{figure}
\hspace*{-0.48\textwidth}
\includegraphics[width=0.48\textwidth]{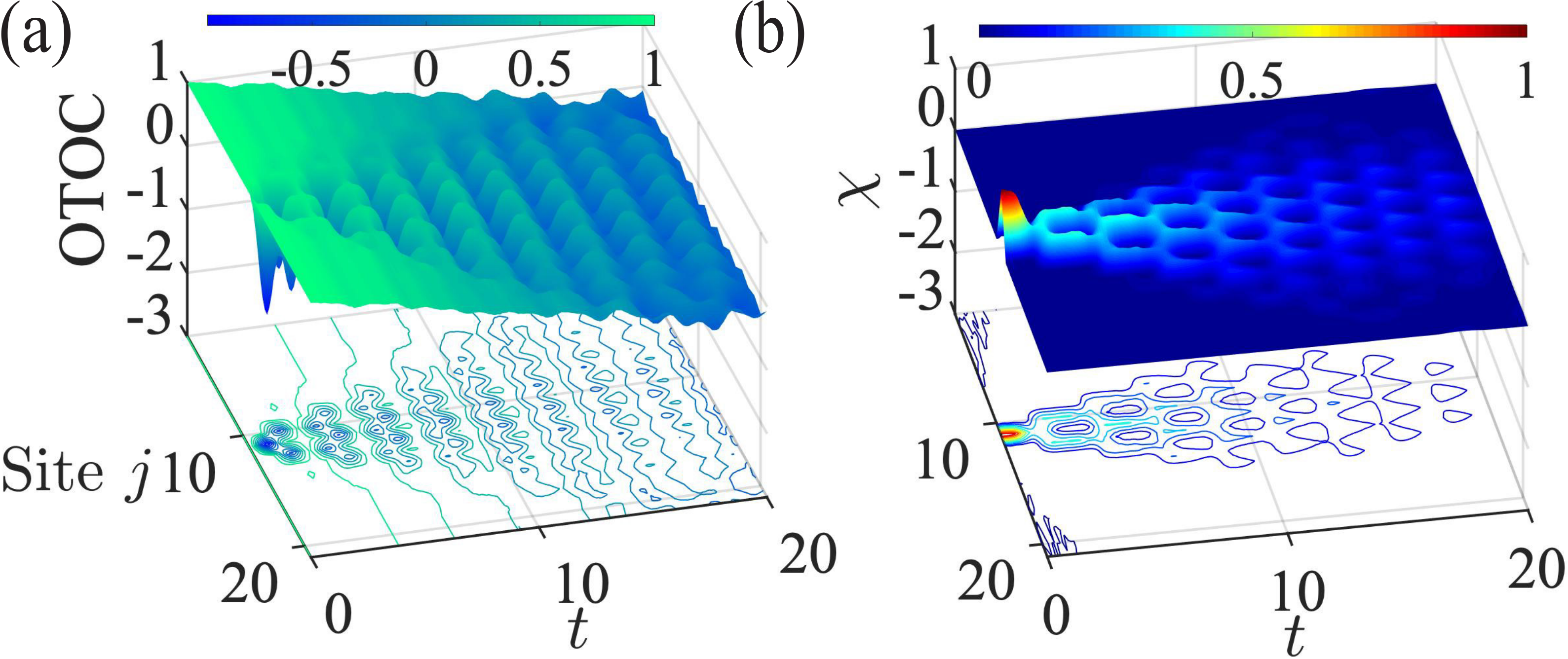} 
\caption{Numerical simulations of the $ZZ$-OTOC (a) and Holevo information (b) dynamics for the Rydberg-atom Hamiltonian approximated by $\mathcal{H}_{\pm}$, with the initial state $\ket{Z_2}$, under experimental parameters $L=21$, $\Omega = 2$, $U_1 =12$, $U_2 = 0.19$ and $\Delta =0.38$ (by ED). Contour lines of the OTOC and Holevo information $\chi$ are projected on the $j-t$ plane.
}
\label{fig:experiment} 
\end{figure}

When approximating the PXP model using the lowest $U(1)$ symmetry sector of the Rydberg-atom Hamiltonian, the  errors mainly come from the invalidity of the condition $U_1 \gg \Omega \gg U_2$. Constrained by the geometry of 1D equally-spaced atoms, $U_1/U_2 \sim 64$ is fixed, such that $\Omega$ around $\sqrt{U_1 U_2}$ can best fulfill the inequality above. Besides, we numerically observe that a small non-zero $\Delta$ in $\mathcal{H}_{\pm}$ can further eliminate the effect of $U_2$, which is probably due to the cancellation of $- \Delta \sum_{i} n_{i}$ and $U_2 \sum_i  n_{i} n_{i+2}$ terms on the $\ket{Z_2}$ initial state. The fact that non-zero detunings amplify the signatures of quantum many-body scars is reminiscent of the recent experiment \cite{Bluvstein2021Controlling} in which periodically driven detunings stabilize the scar revivals, and might be of independent research interest. In Fig. \ref{fig:experiment}, we display the numerical simulations of the $ZZ$-OTOC and Holevo information dynamics for the Rydberg-atom Hamiltonian approximated by $\mathcal{H}_{\pm}$, with the initial state $\ket{Z_2}$, under experimental parameters from \cite{Bernien2017Probing} and an optimized $\Delta$ to increase the oscillation contrast as much as possible (see Appendix. \ref{sec:more numerics} for more details). The linear light cone contour and periodic oscillations inside the light cone can be readily observed.

\section{Conclusions and discussions} 
In summary, we have studied the information scrambling dynamics in quantum many-body scarred systems. We found an unconventional paradigm (a linear light cone with periodic oscillations inside, revealed by OTOCs and Holevo information) intrinsically distinct from the previously studied thermal or MBL systems. Based on perturbation-type calculations in the interaction picture, we provided analytical explanations for this paradigm. In addition, we have also proposed an experiment to measure the predicted exotic information scrambling dynamics with current Rydberg atom technologies. 

Because of the forward and backward Hamiltonian evolution and interleaved local operators, the periodic revivals of OTOCs and Holevo information inside the light cone can not be directly deduced from the eigenstate decomposition of initial states, thus distinguishing our work from the previous literature. For initial states within the scarred subspace, locally encoded information is retrievable elsewhere even at late time, indicating persistent backflow of quantum information and unusual breakdown of quantum chaos.
Our results present an information-theoretic perspective on quantum many-body scars, which would connect to a number of possible directions for future studies, such as Hilbert space fragmentation \cite{Langlett2021Hilbert,Moudgalya2021Quantum,hahn2021information,Khemani2020Localization,buvca2022Out}, the classical OTOC and chaos theory \cite{Ho2019periodic,Michailidis2020Slow,Turner2021Correspondence,Mondal2021Dynamical,sinha2021fingerprint}, robustness of scarred eigenstates under perturbations \cite{Lin2020Slow,Surace2021Exact,Shem2021Fate,Huang2021Stability,Surace2021Quantum}, black hole physics \cite{Bao2017Distinguishability,Guo2018Distinguishing,Qi2022Holevo}, and quantum technology applications including quantum memory and quantum sensing \cite{Serbyn2021quantum,Robust2021Dooley,Lvovsky2009Optical}.

\begin{acknowledgements}
We acknowledge helpful discussions and communications with Xun Gao, Shenglong Xu, and Soonwon Choi. This work is supported by the Frontier Science Center for Quantum Information of the Ministry of Education of China, Tsinghua University Initiative Scientific Research Program, the start-up fund from Tsinghua University, the National Natural Science Foundation of China (Grants No. 12075128 and No. 11905108), and the Shanghai Qi Zhi Institute.
\end{acknowledgements}

\appendix
\section{Details of Analytical Derivations}
\label{sec:analytical}

In this Appendix, we provide detailed analytical derivations of the OTOC dynamics in the early growth region, i.e. Eq. \eqref{Eq:analytical} in the main text, and relevant discussions.
As mentioned in the main text, the essential features of the PXP model can be described by a phenomenological model proposed in \cite{Choi2019emergent}:
\begin{equation}
    H' = H_0' + R = \frac{\Omega}{2} \sum_{i} \sigma_{i}^{x}+\sum_{i} R_{i, i+3} P_{i+1, i+2},
\label{Eq:phen_model}
\end{equation}
where $P_{i,i+1}=(1-\vec{\sigma}_i\cdot \vec{\sigma}_{i+1})/4$ is the projector towards the singlet state of spins $i, i+1$, and $R_{i,j}=\sum_{\mu,\nu}J^{\mu\nu}_{ij}\sigma^\mu_i \sigma^\nu_j$ ($J^{\mu\nu}_{ij}$ are random coupling constants, $\mu,\nu=\{x,y,z\}$). The $H_0'$ and $R$ terms characterize the periodic revivals and thermalization effect in the PXP model respectively. Below we consider a 1D open boundary chain with $L$ $1/2$-spins. 

The $L+1$ scarred eigenstates of $H'$ are all the $x$-direction Dicke states $\ket{s=L/2, S^x=m^x}$, namely the $L+1$ states of the angular momentum $s=L/2$ with $x$-component $m_x=-s, -s+1, \cdots s-1, s$. For instance, $\ket{s=2, S^x=-2} = \ket{----}$, $\ket{s=2, S^x=0} = (\ket{++--}+\ket{+-+-}+\ket{+--+}+\ket{-++-}+\ket{-+-+}+\ket{--++})/\sqrt{6}$, where $\ket{\pm}_i$ are $\pm 1$ eigenstates of $\sigma_i^x$. Since any Dicke state has an explicit Schmidt decomposition:
\begin{align}
\ket{s=\frac{L}{2}, S^x= l & -\frac{L}{2}} = \sum_{n=0}^N \sqrt{P_n(l)} \ket{s=\frac{N}{2}, S^x=n-\frac{N}{2}} \nonumber\\ &\ket{s=\frac{L-N}{2}, S^x=l-n-\frac{L-N}{2}},    
\end{align}
where $P_n(l)=\binom{N}{n} \binom{L-N}{l-n} / \binom{L}{l}$, and $P_{i, i+1} \ket{s=1, S^x=-1,0,1}_{i,i+1} = 0 \quad \forall i$, we have $P_{i, i+1} \ket{s=L/2, S^x=m^x} = 0 \quad \forall i, m_x$. We hence deduce that
\begin{align}
H' \ket{s=\frac{L}{2}, S^x = & m^x} = H_0' \ket{s=\frac{L}{2}, S^x=m^x} \nonumber\\
&= m_x \Omega \ket{s=\frac{L}{2}, S^x=m^x}.
\end{align}

These scarred eigenstates form an exact $su(2)$ algebra (while the scars in the PXP model form an approximate one) \cite{Choi2019emergent}, leading to the periodic revival dynamics governed by the global Rabi oscillation term $H_0'$, from initial states within the scarred subspace (for example, the $z$-direction Dicke states like $\ket{\psi}=\ket{\uparrow\uparrow\cdots\uparrow}$). In contrast, initial states out of the scarred subspace can not be projected out by $P_{i,i+1}$, thus are affected by the $R_{i,i+3}$ terms and have chaotic dynamics.

\begin{figure}
\hspace*{-0.5\textwidth}
\includegraphics[width=0.5\textwidth]{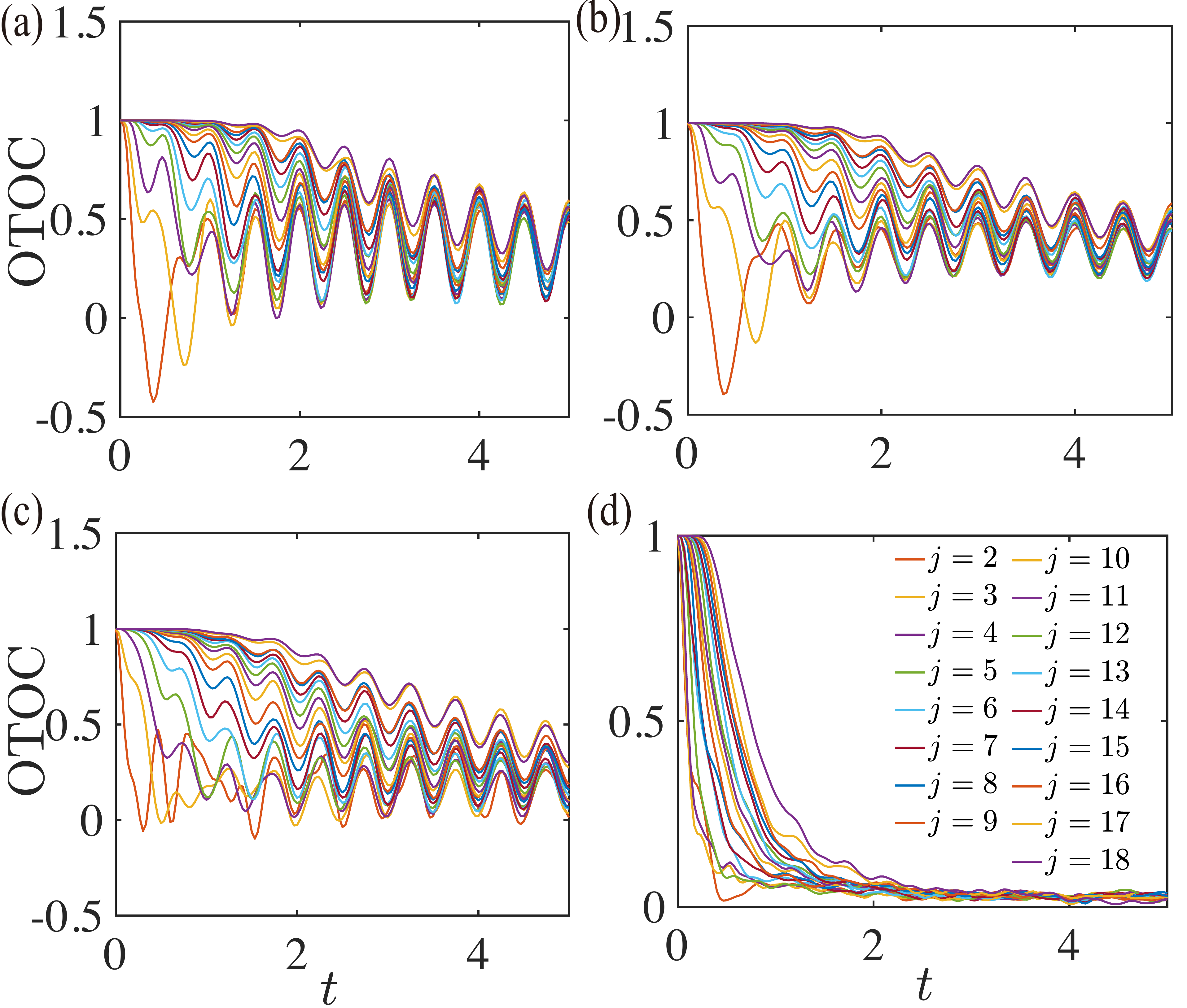} 
\caption{Numerical simulations of the $X Z$-OTOC dynamics (a), (b), (d) and $Z Z$-OTOC dynamics (c) for the $H'$ model, Eq. \eqref{Eq:phen_model}, with the initial states being (a) the $\ket{s=L/2, S^z=-L/2}$ Dicke state, (b) the $\ket{s=L/2, S^z=-L/2+1}$ Dicke state, (c) the $\ket{s=L/2, S^z=0}$ Dicke state and (d) the $Z_2$ N\'eel state ($\ket{\uparrow\downarrow\uparrow\cdots\uparrow\downarrow}$). $i=1$ in $F_{ij}(t)$, $L=18$, calculated by ED.
}
\label{fig:OTOC_toy_SM} 
\end{figure}

The OTOC operator $F_{ij}(t) = \langle \psi | W_i^\dagger V_j^\dagger(t) W_i V_j(t) |\psi\rangle$ can be viewed as the overlap between two time-dependent quantum states $F_{ij}(t) = \langle \psi_2(t) | \psi_1(t) \rangle $:
\begin{align}
&|\psi_1(t) \rangle = W_i e^{i H t} V_j e^{-i H t} |\psi \rangle \nonumber\\
&|\psi_2(t) \rangle =  e^{i H t} V_j e^{-i H t} W_i |\psi \rangle.
\label{Eq:OTOC_overlap}
\end{align}

First of all, for the $ZZ$-OTOC case of the PXP model with the initial state $\ket{Z_2}$ (or the $H'$ model above with the initial state $\ket{\psi}=\ket{\uparrow\uparrow\cdots\uparrow}$), since $\ket{Z_2}$ ($\ket{\psi}$) is the eigenstate of the $\sigma^z_{i(j)}$ operator, the periodic revival dynamics of the quantum state directly give us the periodic and synchronized oscillations of the $ZZ$-OTOC $F_{ij}(t)$. Note that the oscillation period of the $ZZ$-OTOC is supposed to be $T/2=\pi/\Omega$, which is indeed true for the $H'$ model with perfect quantum many-body scars. Because after $T/2$ evolution, $\ket{Z_2}$ evolves to $\ket{Z_2'}=(\prod_{i=1}^L \sigma_i^x) \ket{Z_2}$ ($\ket{\uparrow\uparrow\cdots\uparrow}$ evolves to $\ket{\downarrow\downarrow\cdots\downarrow}$), which is again the eigenstate of the $\sigma^z_{i(j)}$. However, the periodic revival dynamics and emergent $su(2)$ algebra of the PXP model are not perfect \cite{Choi2019emergent}, resulting in smaller peak values of $F_{ij}(t=(2n+1)T/2)$ than those of $F_{ij}(t=n T)$ $(n=0,1,2\cdots)$. Eventually we observe an overall oscillation period $T=2\pi/\Omega$ in numerical simulations. 

Specially for the PXP model, since we consider the dynamics within the constrained Hilbert space (where computational bases with two nearby up spins $\ket{\cdots \uparrow\uparrow \cdots }$ are removed), only the $ZZ$-OTOC dynamics for generic initial states $\ket{\psi}$ are legal, namely within the constrained Hilbert space. The $XZ$-OTOCs for the $\ket{Z_2}$ initial state with $W_i=\sigma_i^x$ acting on an up spin $\ket{\uparrow}_i$ are also legal and have well-defined physical meanings in the overlap interpretation Eq. \eqref{Eq:OTOC_overlap}. Fortunately, the problem of constrained Hilbert space does not appear in the $H'$ model above. Besides, the Holevo information dynamics for general initial states are also not well defined, since it will be difficult to locally encode one bit information on general entangled states, for example the scarred eigenstates.

Now we consider the non-trivial $XZ$-OTOC dynamics of $H'$ with the initial state $\ket{\psi}=\ket{\uparrow\uparrow\cdots\uparrow}$. For $W_1=\sigma^x_1$, $V_r=\sigma^z_r$, $t=n T/2\ (n=0,1,2\cdots)$, according to Eq. \eqref{Eq:OTOC_overlap},
\begin{align}
&\ket{\psi_1(t)}=\sigma^x_1 e^{i H' t} \sigma^z_r e^{-i H' t}\ket{\psi} = (-1)^n \sigma^x_1 \ket{\psi} \nonumber\\
&\ket{\psi_2(t)}=e^{i H' t} \sigma^z_r e^{-i H' t}\sigma^x_1 \ket{\psi},
\end{align}
\begin{equation}
    F(r,t)= \langle \psi_2(t) | \psi_1(t) \rangle =(-1)^n \bra{\phi(t)} \sigma_r^z \ket{\phi(t)}.
\label{Eq:OTOC_ave_form}
\end{equation}
Here by utilizing the property of state recurrence, we successfully convert the four-body OTOC into an observable average on a time-dependent quantum state $\ket{\phi(t)} = e^{-i H' t} \sigma^x_1 \ket{\psi}$. 


In order to further explore the role of the $R$ thermalization term, we adopt the interaction picture of $H_0'=(\Omega/2) \sum_{i} \sigma_{i}^{x}$ to remove the Rabi oscillation effect. We denote the quantum states and operators in the interaction picture with hats:
\begin{equation}
    \ket{\hat{\phi}(t)}=e^{i H_0' t} \ket{\phi(t)} \quad \hat{R}(t) = e^{i H_0' t} R e^{-i H_0' t}.
\end{equation}
The quantum state evolution in the $H_0'$ interaction picture is 
\begin{equation}
    i\partial_t \ket{\hat{\phi}(t)} = \hat{R}(t) \ket{\hat{\phi}(t)},
\end{equation}
\begin{align}
&\ket{\hat{\phi}(t)} = \hat{U}(t)\ket{\hat{\phi}(0)}=\hat{U}(t) \sigma_1^x\ket{\psi} \nonumber\\
&\hat{U}(t) = T_t \left(\exp(-i\int_0^{t} d t' \hat{R}(t'))\right),
\label{Eq:state_evol_int}
\end{align}
where we have introduced the time ordering operator $T_t$. According to Eq. \eqref{Eq:OTOC_ave_form}, at the special time points $t = n T/2$, the OTOCs in the interaction picture have the expression
\begin{align}
F(r,t) &= (-1)^n \bra{\hat{\phi}(t)} e^{i H_0' t} \sigma_r^z e^{-i H_0' t} \ket{\hat{\phi}(t)} \nonumber\\
    &= (-1)^n \bra{\hat{\phi}(t)} e^{i \Omega \sigma_r^x t / 2} \sigma_r^z e^{-i \Omega \sigma_r^x t / 2} \ket{\hat{\phi}(t)} \nonumber\\
    &= \bra{\hat{\phi}(t)} \sigma_r^z \ket{\hat{\phi}(t)}.
\label{Eq:OTOC_int_pic}
\end{align}

\begin{figure*}
\hspace*{-0.95\textwidth}
\includegraphics[width=0.95\textwidth]{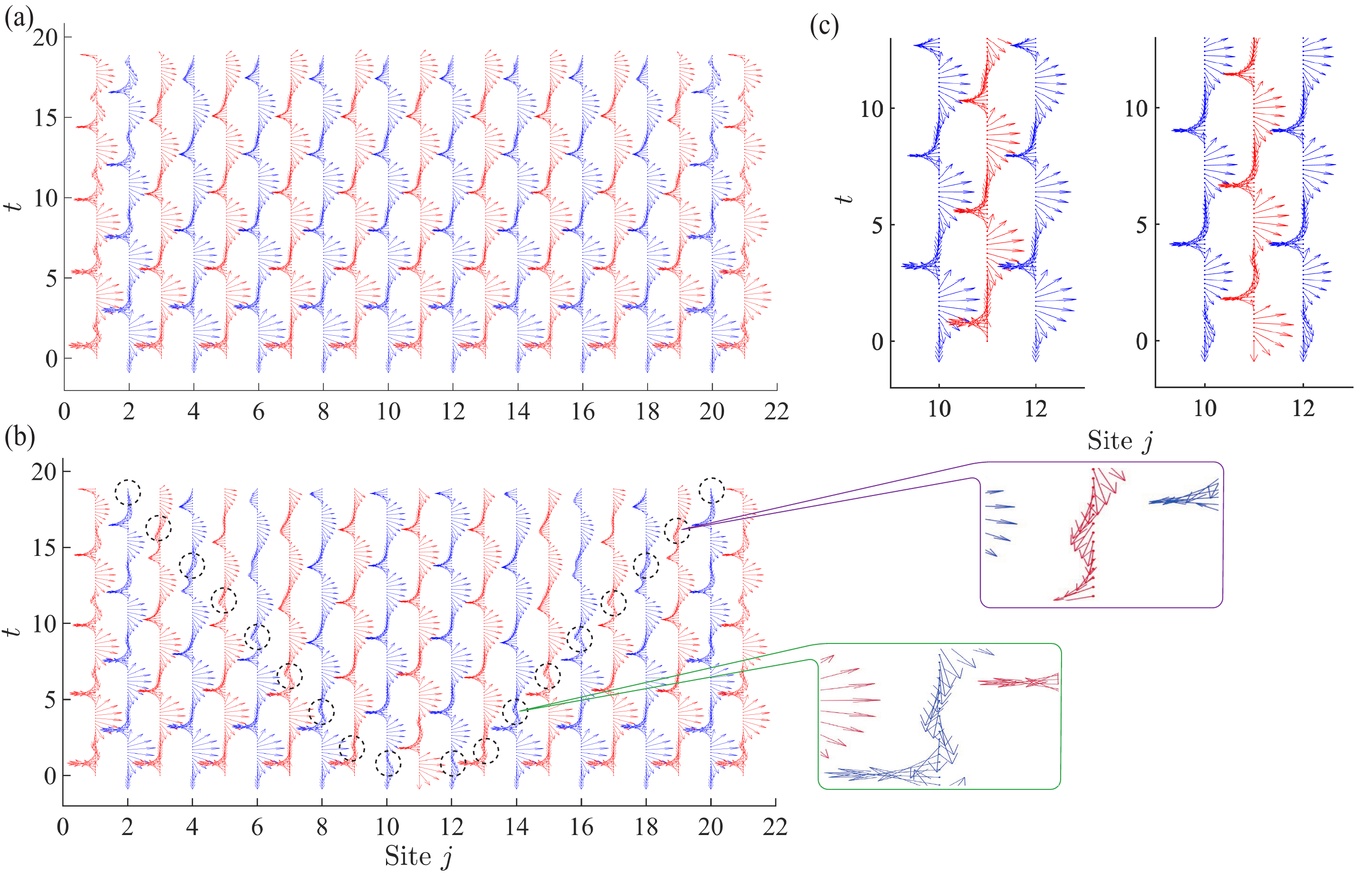} 
\caption{Kinetically constrained spin rotations of the PXP model and the relationship with Holevo information dynamics. The rotation dynamics of Bloch vectors $\vec{a}(t)$ with the initial state (a) $|Z_2 \rangle$ and (b) $\sigma_{\lceil L/2 \rceil}^x\ket{Z_2}$, $L=21$. The dashed circles in plot (b) indicate the inverse rotation regions, which match up with the contour of the linear light cone. The green and purple boxes are enlarged images for these regions. Plot (c) focuses on rotation dynamics for the three middle spins of $|Z_2 \rangle$ (left) and $\sigma_{\lceil L/2 \rceil}^x\ket{Z_2}$ (right). Due to the kinetic constraints, the retarded spin rotations lead to the periodic distinguishabilty of single-qubit reduced density matrices $\rho_j(t)$ and $\rho^\prime_j(t)$. The Holevo information $\chi_j(t)$ hence forms the persistent oscillation pattern inside the light cone.}
\label{fig:vec_rot} 
\end{figure*}

Below we consider the so-called early growth region of the OTOC, namely the time interval from $t=0$ to the time $t$ just before the wavefront of heat flow arrives at the site $r$. Then we can split the evolution time $t$ into $r$ pieces with $\Delta t=t/r,\  J\Delta t \ll 1$ ($J$ is the average energy scale of all the $J^{\mu\nu}_{ij}$), to treat the dynamics perturbatively:
\begin{align}
    \hat{U}(t) =& T_t \left(\exp(-i\int_0^{t} d t' \hat{R}(t'))\right) \nonumber\\
    =& \prod_{n=0}^{r-1} \exp(-i\int_{n\Delta t}^{(n+1)\Delta t} d t' \hat{R}(t')) \nonumber\\
    \approx& \prod_{n=0}^{r-1} (1 - i \hat{R}(n\Delta t)\Delta t),
\label{Eq:r_terms_prod}
\end{align}
where 
\begin{align}
&\hat{R}(n\Delta t) = \sum_i \hat{R}^i (n \Delta t) \nonumber\\
&= \sum_i \left( e^{i H_0' n\Delta t} R_{i, i+3} P_{i+1, i+2} e^{-i H_0' n\Delta t} \right).
\end{align}

According to Eq. \eqref{Eq:state_evol_int} and \eqref{Eq:OTOC_int_pic}, besides the leading ``$1$" term in Eq. \eqref{Eq:r_terms_prod}, the OTOC $F(r,t=n T/2)$ will be dominantly influenced by the following operator product series:
\begin{equation}
    (-i)^r (\Delta t)^r \hat{R}^{r}\left( (r-1)\Delta t\right) \cdots \hat{R}^2(\Delta t) \hat{R}^1(0) \sigma^x_1 \ket{\psi}.
\label{Eq:ope_series}
\end{equation}
The operator product series Eq. \eqref{Eq:ope_series} vividly characterizes propagation of the thermalization effect from the site $1$ to site $r$. Since $H_0'$ only contains single-body operators, other $\hat{R}^{i}( n\Delta t)$ operator product series acting on $\sigma^x_1 \ket{\psi}$ will either be zero due to the projectors $P_{i,i+1}$, or unable to reach the site $r$ and affect the perfect revival dynamics there. Inserting Eq. \eqref{Eq:ope_series} into Eq. \eqref{Eq:OTOC_int_pic}, we obtain that the leading correction term of the OTOC $F(r,t=n T/2)$ can be bounded by
\begin{align}
&||[ \sigma_r^z, \hat{R}^{r}\left( (r-1)\Delta t\right) \hat{R}^{r}\left( (r-2)\Delta t\right) \cdots \nonumber\\
&\hat{R}^2(\Delta t) \hat{R}^1(0) ]_{\pm}|| (\Delta t)^r \leq (a J \Delta t)^r,
\label{Eq:corr_term}
\end{align}
where $a$ is some model-dependent $O(1)$ constant, $[\cdot]_{\pm}$ denotes commutator($-$) and anti-commutator($+$) (depending on the parity of $r$ to choose which), and $||\cdot||$ denotes the operator norm. The bound in Eq. \eqref{Eq:corr_term} does not depend on $T$ or $\Omega$ (which stands for the resolution in time), so we are able to generalize the special time points $t=n T/2$ to arbitrary $t$ before the wavefront arrives. Finally we have the following OTOC behaviors in the early growth region:
\begin{equation}
    F(r,t)\sim 1 - (\frac{a J t}{r})^r.
\label{Eq:OTOC_final_form}
\end{equation}

Eq. \eqref{Eq:OTOC_final_form} readily depicts a linear light cone structure $t \propto r/J$ in the early growth region. While we specifically compute the $X Z$-OTOC of the phenomenological model Eq. \eqref{Eq:phen_model}, the physical pictures of local thermalization and ballistic propagation hold for other quantum many-body scarred systems, other OTOCs and Holevo information dynamics, and can be applied to explain the linear light cone structure observed in numerical simulations. An additional remark is that the perturbation calculations above are similar to the derivations of Kubo formula in the linear response theory, and the correction term Eq. \eqref{Eq:ope_series} corresponds to the $r$-th order response to the perturbation $\hat{R}(t)$ \cite{Bruus2004Many}.

Numerical simulations for the OTOC dynamics of the $H'$ model are displayed in Fig. \ref{fig:OTOC_toy_SM}. The persistent and synchronized oscillations are readily observed for initial states within the scarred subspace in plots (a), (b) and (c). In contrast for the $Z_2$ N\'eel state, OTOCs constantly decay to zero without any revival. These numerical results further confirm the generality of our conclusions to other models with quantum many-body scars. In future studies it is also interesting to explore whether the formation mechanisms of quantum many-body scars \cite{Serbyn2021quantum} will affect the information scrambling dynamics in corresponding models.

\begin{figure}
\hspace*{-0.5\textwidth}
\includegraphics[width=0.5\textwidth]{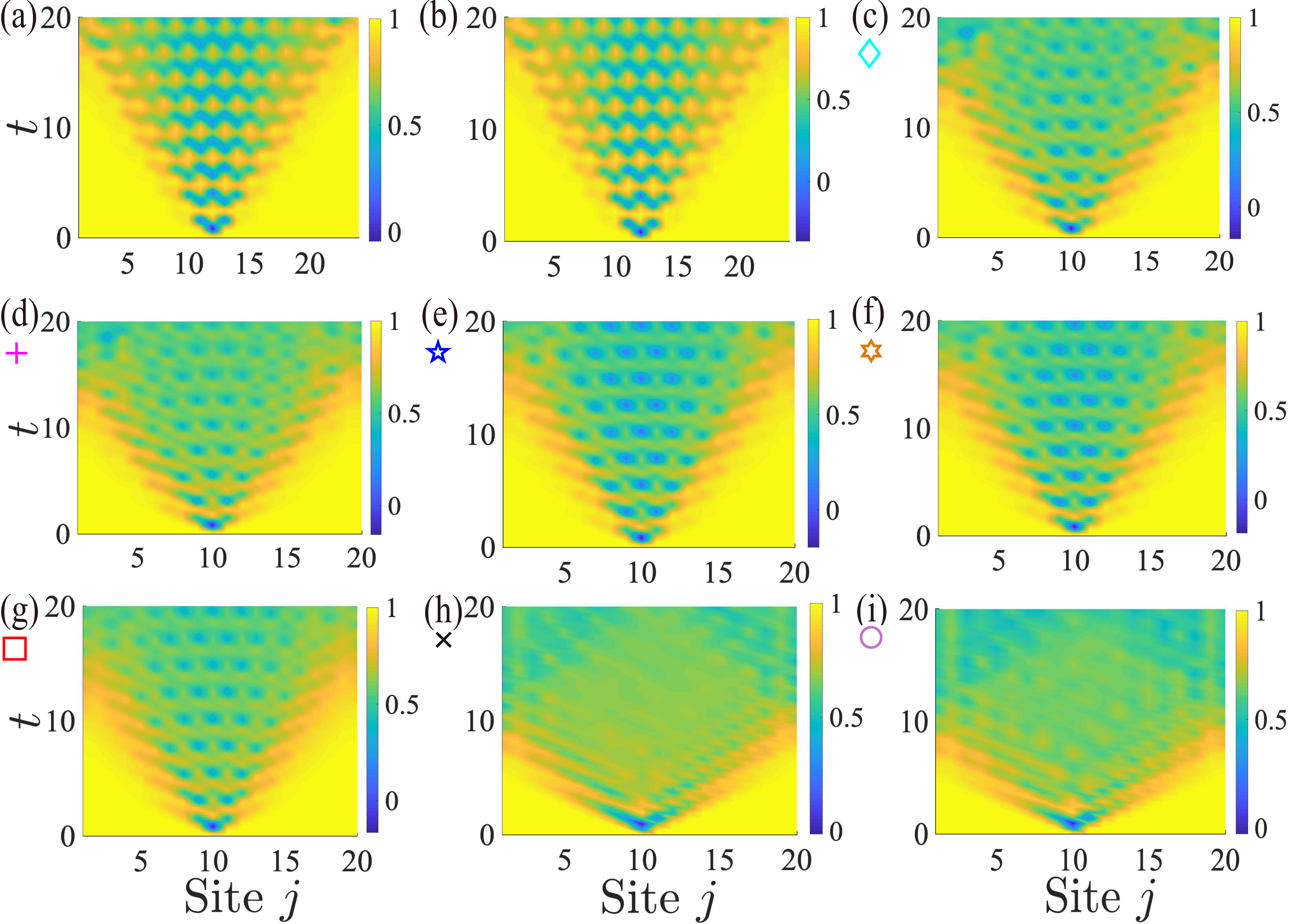} 
\caption{Numerical simulations of $Z Z$-OTOC dynamics for the $L=20$ PXP model in the open boundary condition, with initial states being (a) $\frac{1}{\sqrt{2}}(|Z_2\rangle + |Z_2'\rangle) $, $|Z_2'\rangle = \prod_i \sigma_i^x |Z_2\rangle$; (b) $\sqrt{\frac{2}{3}} |Z_2\rangle + \sqrt{\frac{1}{3}}|Z_2'\rangle $; (c)-(f) scarred and thermal energy eigenstates, marked by the same labels in Fig. \ref{fig:overlap}.
}
\label{fig:otherscarZZ} 
\end{figure}

\begin{figure}
\hspace*{-0.47\textwidth}
\includegraphics[width=0.47\textwidth]{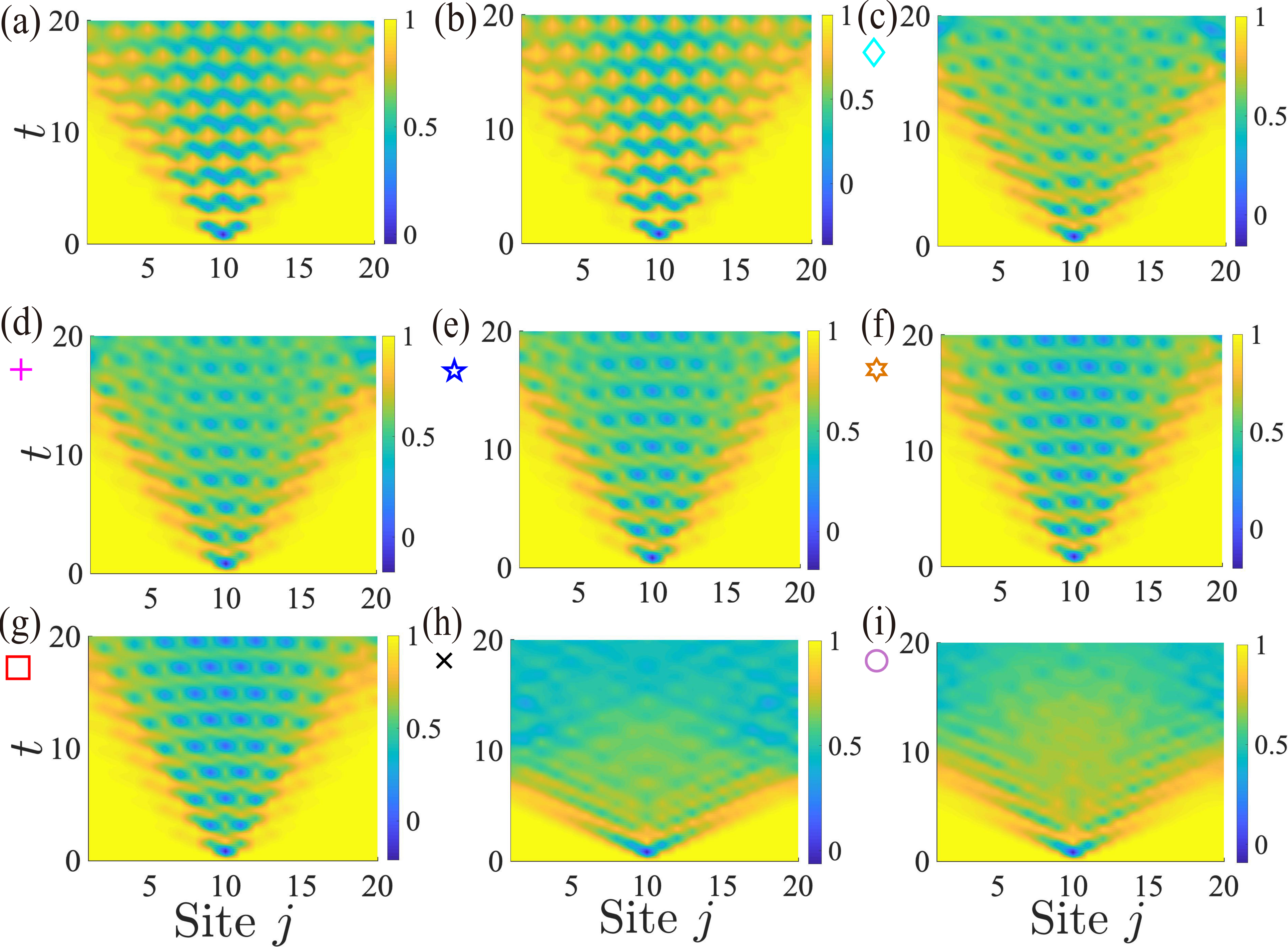} 
\caption{Numerical simulations of $Z Z$-OTOC dynamics for the $L=20$ PXP model in the periodic boundary condition, with initial states being (a) $\frac{1}{\sqrt{2}}(|Z_2\rangle + |Z_2'\rangle) $, $|Z_2'\rangle = \prod_i \sigma_i^x |Z_2\rangle$; (b) $\sqrt{\frac{2}{3}} |Z_2\rangle + \sqrt{\frac{1}{3}}|Z_2'\rangle $; (c)-(f) scarred and thermal energy eigenstates, marked by the same labels in Fig. \ref{fig:overlapclose}.
}
\label{fig:otherscarZZclose} 
\end{figure}

As mentioned in the main text, the persistent and synchronized oscillations inside the light cone still appear even if the initial states are replaced with scarred eigenstates or superposition states of $\ket{Z_2}$ and $\ket{Z_2'}=(\prod_{i=1}^L \sigma_i^x) \ket{Z_2}$ (see Fig. \ref{fig:otherscarZZ} and \ref{fig:otherscarZZclose}), which confirms that periodic revivals of OTOCs are general phenomena for initial states within the scarred subspace, not some fine-tuned results. We have attributed the unusual revival dynamics inside the light cone to the local rather than global thermalization and robustness of scarred eigenstates against local perturbations. In this sense, the periodic revival behaviors of OTOCs in quantum many-body scarred systems are somewhat similar to the dynamics in time crystals \cite{Sacha2017Time,Else2020Discrete,Huang2018Clean,buvca2022Out}: The $Z Z$-OTOC dynamics for the $\ket{Z_2}$ initial state is an analogue to the time crystals without perturbations; while the $X Z$-OTOCs for $\ket{Z_2}$ and $Z Z$-OTOCs for general initial states within the scarred subspace correspond to the perturbed time crystals, which will exhibit certain robustness and maintain the synchronized oscillations. This topic might need deeper understanding and more powerful techniques to deal with, so that is expected to inspire more analytical studies in the future. One possible direction is about the relation with the classical OTOC and chaos theory \cite{Ho2019periodic,Michailidis2020Slow,Turner2021Correspondence,Mondal2021Dynamical,sinha2021fingerprint}. According to the quantum-classical correspondence principle, we may replace the quantum commutators in the OTOC with classical Poisson brackets. By utilizing the semi-classical Lagrangian of quantum many-body scarred systems, like the one proposed in \cite{Ho2019periodic} with the time-dependent variational principle (TDVP), we are able to compute the Poisson brackets ($\{\cdot\}_{P.B.}$) of some observables like $\{\cos\theta_i(t), \cos\theta_j(0)\}_{P.B.}\sim \sin\theta_i(t) \pardif{\theta_i(t)}{\theta_i(0)}$ ($\theta$ is the polar angle on the spin-1/2 Bloch sphere). The oscillating dynamics of $\sin^2 \theta_i(t)$ terms from some special initial states like $\ket{Z_2}$ will roughly lead to the periodic revivals of classical OTOCs.

\begin{figure}
\hspace*{-0.49\textwidth}
\includegraphics[width=0.49\textwidth]{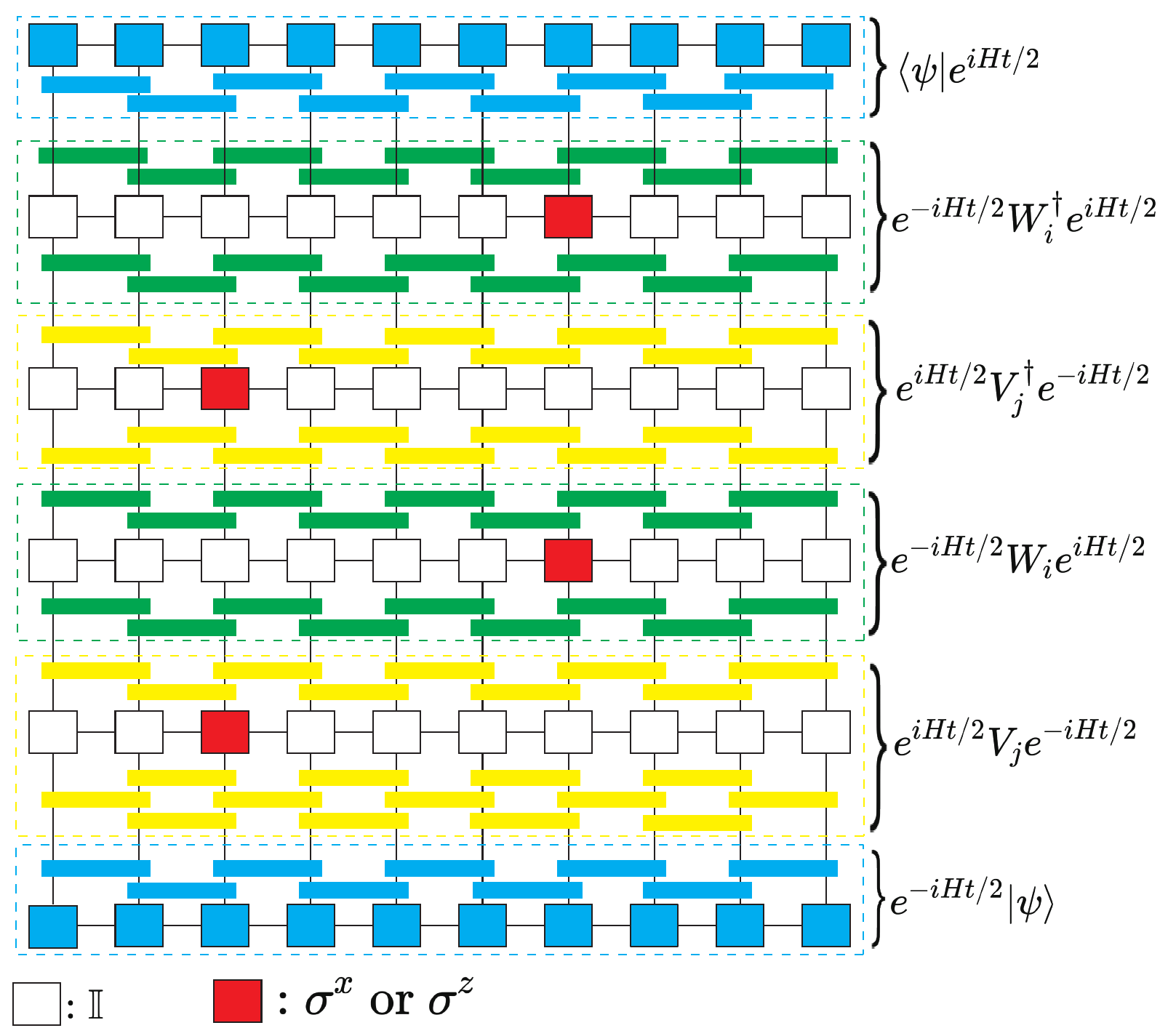} 
\caption{The time-splitting MPO algorithm for OTOC calculations. The blue squares denote MPS tensors and the white (red) squares denote MPO tensors. All the rectangles stand for the Trotterized Hamiltonian evolution blocks. The total evolution time $t$ has been equally split to each part of $F_{ij}(t) = \langle \psi(t/2) | W_i^\dagger(-t/2) V_j^\dagger(t/2) W_i(-t/2) V_j(t/2) |\psi(t/2) \rangle$ to reduce the required bond dimension. The overall contraction of the tensor network above will give the OTOC $F_{ij}(t)$.}
\label{fig:mpo} 
\end{figure}

\section{More about Holevo Information Dynamics}
\label{sec:Holevo}


As mentioned in the main text, we regard the reduced Hamiltonian evolution on subsystems as quantum communication channels in the original setup of Holevo information \cite{Holevo1973bounds} and use it to study the information scrambling dynamics. Through two sets of Hamiltonian evolution on different initial states, $\ket{\psi}$ and $\sigma^x_i \ket{\psi}$, we demonstrate how one-bit local information at the site $i$ scrambles into the entire system by computing the Holevo information on the site $j$
$\chi_j(t) = S\left(  (\rho_j(t) +\rho'_j(t))/2 \right) -  \left( S( \rho_j(t) ) + S(\rho'_j(t) )\right)/2,$ where $\rho_j(t)$ and $\rho^\prime_j(t)$ are reduced density matrices of the $j$-th spin after the Hamiltonian evolution for the initial state $\ket{\psi}$ and $\sigma^x_i \ket{\psi}$ respectively. $\chi_j(t)$ measures how much information one could obtain by any local probe on the $j$-th site for these two sets of evolution. If the many-body dynamics follow the predictions of ETH and the computational bases $\ket{\psi}$ and $\sigma^x_i \ket{\psi}$ bear the same energy with respect to the Hamiltonian, at the late time, local single-body density matrices $\rho_j(t)$ and $\rho^\prime_j(t)$ will become indistinguishable. The Holevo information $\chi_j(t)$ will quickly peak when the effect of site $i$ arrives and diminish close to zero afterwards (so does the case of $\ket{\bm{0}}$ and $\sigma_{\lceil L/2 \rceil}^x\ket{\bm{0}}$). In contrast, we have observed persistent and synchronized oscillations inside the light cone for the case of $\ket{Z_2}$ and $\sigma_{\lceil L/2 \rceil}^x \ket{Z_2}$.

When unpacking the dynamics of Holevo information, we find that the kinetic constraints of the PXP model play an important role for the periodic oscillations inside the light cone: According to the PXP Hamiltonian, each spin will rotate freely around the $x$ axis if both its neighbors are in the $\ket{\downarrow}$ state; otherwise its rotation dynamics are frozen. In Fig. \ref{fig:vec_rot}, we display all the single-qubit reduced density matrices used for the computation of Holevo information. The density matrix of a spin-1/2 can be written as $\rho = (I+ \vec{a}\cdot \vec{\sigma})/2$, where $\vec{\sigma}=(\sigma_x, \sigma_y, \sigma_z)$, and $\vec{a}$ is the Bloch vector on the Bloch sphere, such that the dynamics of $\rho(t)$ can be fully parametrized by $\vec{a}(t)$. For the PXP Hamiltonian and initial states being the $z$-direction computational bases, $a_x(t)\equiv 0$. We plot the rotation dynamics of $(a_y(t), a_z(t))$ in Fig. \ref{fig:vec_rot}(a), (b) with initial states being $|Z_2 \rangle$ and $\sigma_{\lceil L/2 \rceil}^x\ket{Z_2}$ respectively. Note that the norm of $\vec{a}(t)$ is not a constant number due to the entanglement between different spins. We can readily observe that the rotation speeds of spins are not constant, which are determined by the orientations of neighbor spins.

In plot (a), for the initial state $|Z_2 \rangle$, all the spins rotate anti-clockwise and the spin configuration has a two-site translational symmetry (indicated by red and blue colors). On the contrary, in plot (b) for the initial state $\sigma_{\lceil L/2 \rceil}^x\ket{Z_2}$, we find some strange inverse rotation regions (spins rotate clockwise for a short period). Interestingly, the positions of inverse rotation regions match up with the contour of the linear light cone to a good accuracy. These strange inverse rotations are a direct result of the kinetic constraints: For the three middle spins of $\sigma_{\lceil L/2 \rceil}^x\ket{Z_2}$, $\ket{\cdots \uparrow \downarrow \downarrow \downarrow \uparrow \cdots}$, they are forbidden to simultaneously rotate to the $\ket{\uparrow}$ state, resulting in the retarded rotations of the $j=10$ and $j=12$ spins (see plot (c) for a zoom-in image). Moreover, this retarded rotation behavior propagates outwards like a row of dominoes, forming a series of inverse rotation regions indicated by the dashed circles. The physical picture is as follows: On the anti-ferromagnetic background $\ket{Z_2}$, we locally create a domain wall in the middle of the spin chain $\ket{\cdots \uparrow \downarrow \downarrow \downarrow \uparrow \cdots}$, the quasi-particle travels ballistically in the following quench dynamics \cite{Jurcevic2014Quasiparticle}, which is consistent with the local thermalization picture discussed in the previous section. 

In plot (b), after the inverse rotation regions, the anti-clockwise kinetically constrained rotations are no more affected, and the two-site translational symmetry is restored inside the light cone. Because of the retarded rotations, the spin orientations inside the light cone become \textit{periodically} distinguishable for the $|Z_2 \rangle$ and $\sigma_{\lceil L/2 \rceil}^x\ket{Z_2}$ cases, leading to the periodic oscillation pattern of Holevo information (compare Fig. \ref{fig:vec_rot}(c) and Fig. \ref{fig:numerics}(e)). The analyses above once again confirm the conclusion that: The unusual information revival dynamics inside the light cone are due to the local rather than global thermalization and robustness of scarred eigenstates against local perturbations. After the wavefront of thermalization effect passes, the kinetically constrained rotations are preserved, causing the persistent oscillations of Holevo information inside the light cone.

\begin{figure}
\hspace*{-0.49\textwidth}
\includegraphics[width=0.49\textwidth]{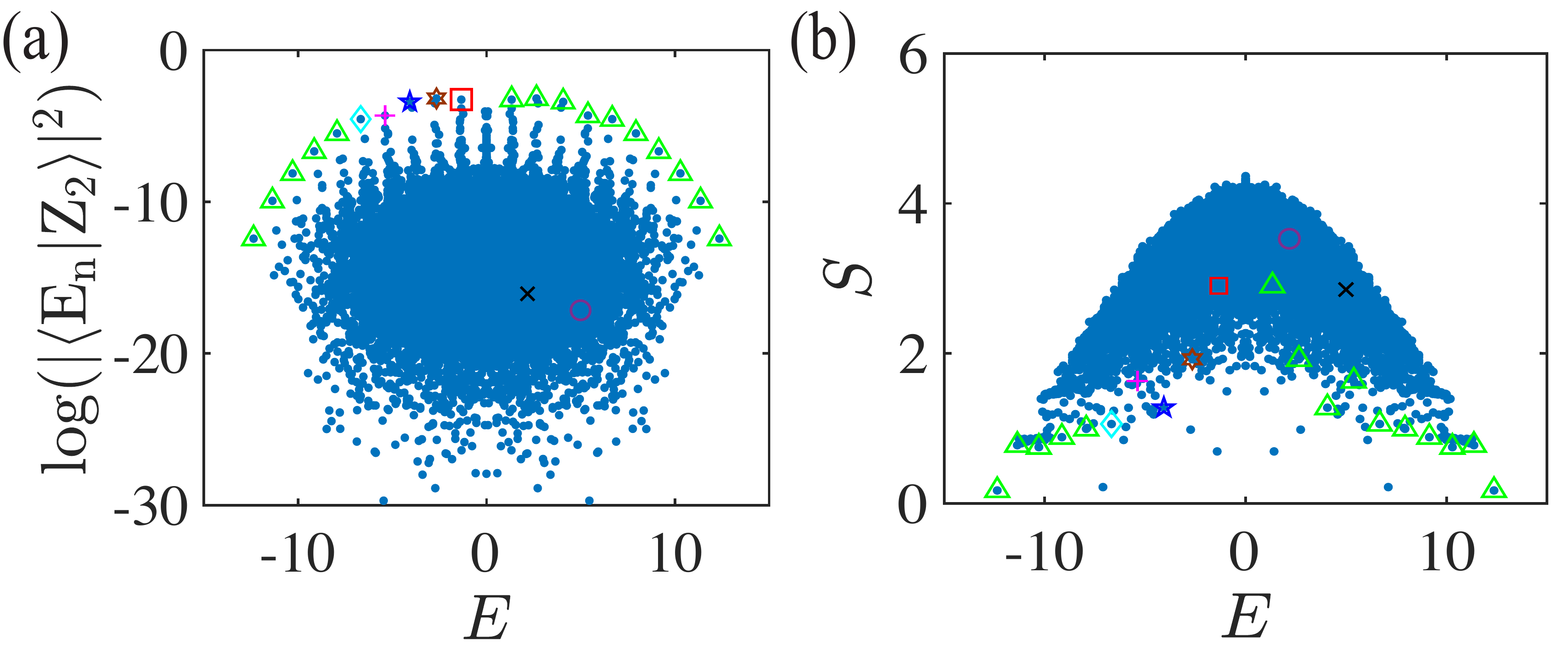} 
\caption{(a) Overlap between the $|Z_2\rangle$ state and each energy eigenstate $|E_n\rangle$ of a $L=20$ PXP model with the open boundary condition. (b) The half-chain entanglement entropy $S$ for each energy eigenstate. The scarred eigenstates form a special band in the top of plot (a). The scars together with two thermal eigenstates in the bulk are marked by labels consistent with those in Fig. \ref{fig:otherscarZZ}.}
\label{fig:overlap} 
\end{figure}

\begin{figure}
\hspace*{-0.49\textwidth}
\includegraphics[width=0.49\textwidth]{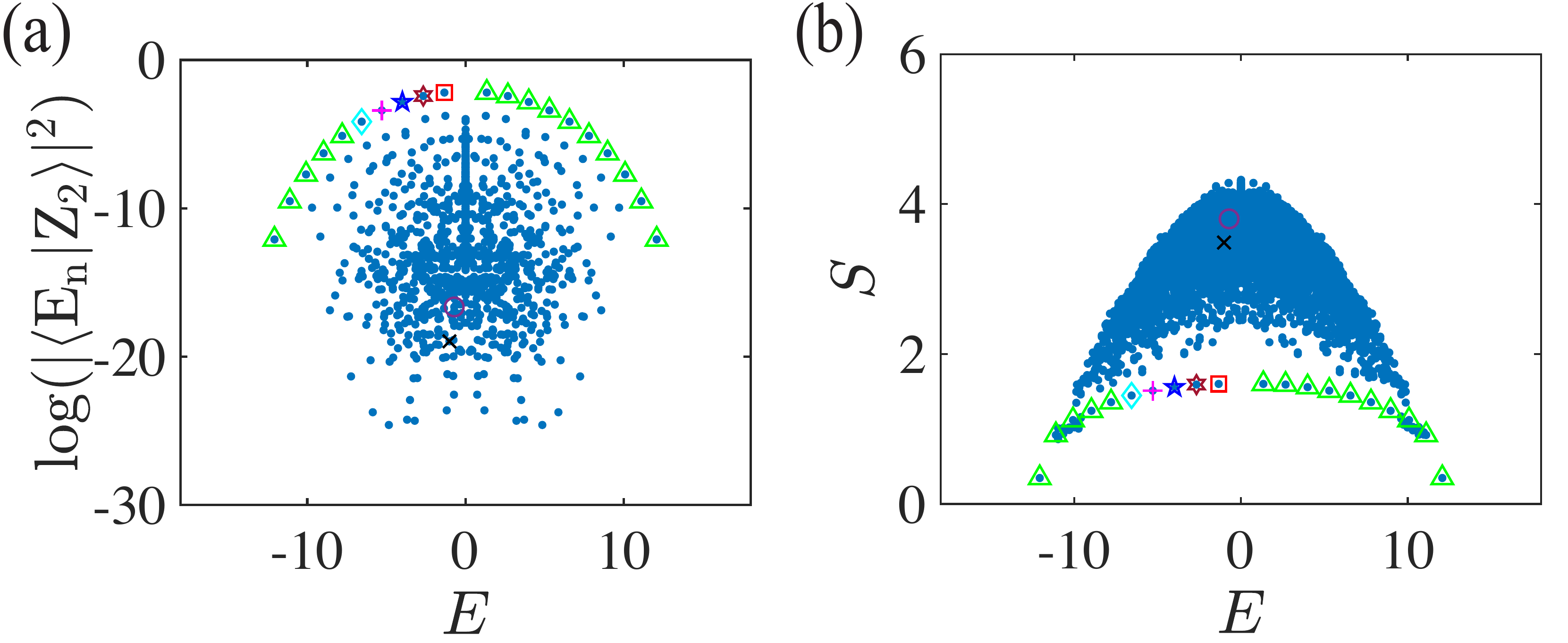} 
\caption{(a) Overlap between the $|Z_2\rangle$ state and each energy eigenstate $|E_n\rangle$ of a $L=20$ PXP model with the periodic boundary condition (eigenstates with log-scaled overlap less than $-30$ are not shown to fit the plotting range). (b) The half-chain entanglement entropy $S$ for each energy eigenstate. The scarred eigenstates form a special band in the top of plot (a). The scars together with two thermal eigenstates in the bulk are marked by labels consistent with those in Fig. \ref{fig:otherscarZZclose}.}
\label{fig:overlapclose} 
\end{figure}

\begin{figure*}
\hspace*{-0.8\textwidth}
\includegraphics[width=0.8\textwidth]{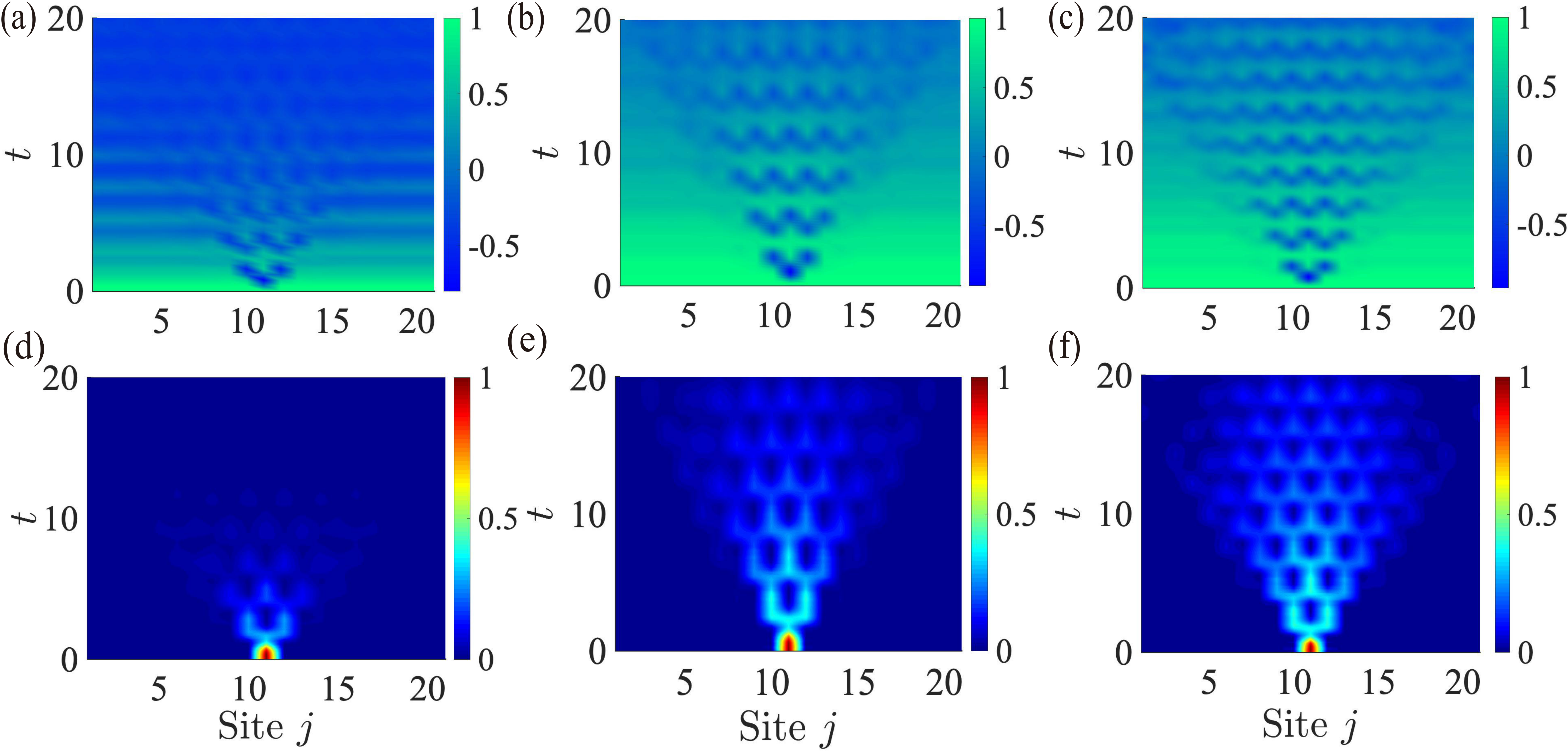} 
\caption{Numerical simulations of the $Z Z$-OTOC dynamics (a)-(c) and Holevo information dynamics (d)-(f) for the experimental Rydberg-atom Hamiltonian, with the initial state $\ket{Z_2}$ and experimental parameters: (a), (d) $L=21$, $\Omega =2$, $U_1 =24$, $U_2 = 0.38$ and $\Delta =0$; (b), (e) $L=21$, $\Omega =1.5$, $U_1 =12$, $U_2 = 0.19$ and $\Delta =0.19$; (c), (f) $L=21$, $\Omega =2$, $U_1 =12$, $U_2 = 0.19$ and $\Delta =0.38$.
}
\label{fig:otherscarZZexp} 
\end{figure*}

\section{Numerical Methods and More Results}
\label{sec:more numerics}

In this section, we illustrate the numerical methods used in the main text and provide more numerical results of the OTOC and Holevo information dynamics in quantum many-body scarred systems. We numerically simulate the $Z Z$- and $X Z$- OTOC dynamics of the PXP model using the time-splitting matrix product operator (MPO) method \cite{xu2020accessing,Zhang2021Anomalous} up to system size $L=41$, with the maximum bond dimension 300 and a Trotter step $d t = 0.05$. Fig. \ref{fig:mpo} displays a pictorial illustration of the algorithm. Specifically, we rewrite the OTOC $F_{ij}(t) = \langle \psi | W_i^\dagger V_j^\dagger(t) W_i V_j(t) |\psi\rangle$  as
\begin{align}
F_{ij}(t) &= \langle \psi(t/2) | W_i^\dagger(-t/2) V_j^\dagger(t/2) \nonumber\\
&W_i(-t/2) V_j(t/2) |\psi(t/2) \rangle,   
\end{align}
where $|\psi(t/2) \rangle = e^{-i H t/2} |\psi \rangle$. Usually in order to maintain the simulation accuracy, we need to increase the bond dimension when extending the evolution time $t$, because of the increase of entanglement entropy. Here by equally splitting the Hamiltonian evolution to each part of $F_{ij}(t)$, we are able to reduce the support of the scrambled operators $V_j(t)$. For a fixed evolution time, this algorithm leads to notable reduction of the required bond dimension compared with previous MPO algorithms \cite{schollwock2011density,xu2020accessing}. 

We use the time-evolving block decimation (TEBD) algorithm \cite{vidal2003efficient,schollwock2011density} based on the matrix product state (MPS) ansatz to simulate the Holevo information dynamics for the PXP model up to system size $L=41$, with the maximum bond dimension $100$ and a Trotter step $d t = 0.05$. The relatively low entanglement entropy of scarred eigenstates [Fig. \ref{fig:overlap}(b) and \ref{fig:overlapclose}(b)] makes the simulations for the $\ket{Z_2}$ initial state much more efficient, yet this advantage no more exists for the $\ket{\bm{0}}$ case. All the MPO and MPS based numerical simulations are carried out with the ITensor library \cite{fishman2020itensor}. 

Numerical calculations for small system size ($L\sim 20$) are performed with the exact diagonalization (ED) method in the constrained Hilbert space of the PXP model, for instance, Fig. \ref{fig:numerics}(f) and Fig. \ref{fig:experiment}.

In the main text, we have shown that the OTOC dynamics exhibit exotic periodic revivals inside the light cone for the initial state $|Z_2 \rangle$. In Fig. \ref{fig:otherscarZZ} and Fig. \ref{fig:otherscarZZclose}, we present the $ZZ$-OTOC dynamics of the PXP model for more general initial states in the open boundary condition (with boundary terms $\sigma^x_1 P_2$ and $P_{L-1} \sigma^x_L$) and periodic boundary condition, respectively. 
First the $Z Z$-OTOC dynamics for superposition states of $|Z_2\rangle$ and $|Z_2'\rangle = \prod_i \sigma_i^x |Z_2\rangle$ are displayed in Fig. \ref{fig:otherscarZZ}(a), (b). Second, in Fig. \ref{fig:otherscarZZ}(c)-(i), we take the energy eigenstates of the PXP Hamiltonian as initial states for the $Z Z$-OTOC calculations. The eigenstates are marked with the corresponding labels in Fig. \ref{fig:overlap}. We observe that despite lower oscillation contrast compared to the $\ket{Z_2}$ case, the pattern of persistent and synchronized oscillations for $Z Z$-OTOCs still exists for the superposition states of $|Z_2\rangle$ and $|Z_2'\rangle$ (a)-(b), scarred energy eigenstates (c)-(g), and also general superposition states of scarred eigenstates. However, the oscillation pattern is absent for typical thermal eigenstates (h)-(i). The numerical results indicate that the periodic revivals of OTOCs inside the light cone are not some fine-tuned results for the initial state $|Z_2\rangle$, but a general phenomenon for a certain class of states within the non-thermal scarred subspace. The information scrambling dynamics for quantum many-body scarred systems are intrinsically different from the thermal or many-body localized systems. One additional remark is that we have calculated the $Z Z$-OTOC dynamics for all the scarred eigenstates marked in Fig. \ref{fig:overlap} and found that the periodic oscillation pattern always exists while the oscillation contrast fades away for a few scarred eigenstates near the ground state, which is probably due to the finite size effect.

In order to distinguish the scarred eigenstates and typical thermal eigenstates, we show the overlap between the $|Z_2\rangle$ state and each energy eigenstate $|E_n\rangle$ of the PXP Hamiltonian with the open boundary condition in Fig. \ref{fig:overlap}(a), and the half-chain entanglement entropy $S$ of energy eigenstates in Fig. \ref{fig:overlap}(b) \cite{Turner2018weak,Turner2018quantum}. All the scarred eigenstates and two thermal eigenstates in the bulk are marked by labels consistent with those in Fig. \ref{fig:otherscarZZ}. The half-chain entanglement entropy of two scarred eigenstates near $E=0$ is relatively larger compared with other scarred eigenstates, which is probably due to the open boundary condition. In order to rule out the possible effects of boundary conditions, we display the corresponding results of the PXP model with the periodic boundary condition in Fig. \ref{fig:otherscarZZclose} and Fig. \ref{fig:overlapclose}. Compared with Fig. \ref{fig:otherscarZZ} and Fig. \ref{fig:overlap}, these results do not show distinct differences despite different boundary conditions.

We present more numerical simulations of $Z Z$-OTOC and Holevo information dynamics of the experimental Rydberg-atom Hamiltonian (Eq. \eqref{Eq:RydbergHam} and related $\mathcal{H}_{\pm}$ in the main text) in Fig. \ref{fig:otherscarZZexp} to demonstrate their measurable signatures. In plots (a), (d), the experimental parameters are the same as those used in \cite{Bernien2017Probing}, where $\Omega =2$, $U_1 =24$, $U_2 = 0.38$ and $\Delta =0$. We observe that the periodic oscillations display a much lower contrast than that of the PXP model, which is induced by the invalidity of the condition $U_1 \gg \Omega \gg U_2$. Constrained by the geometry of 1D equally-spaced atoms, $U_1 /U_2 \sim 64$ is fixed, such that $\Omega$ around $\sqrt{U_1 U_2}$ can best fulfill the condition above. 
With this motivation, we show the results in plots (b), (e), with parameters $\Omega =1.5$, $U_1 =12$, $U_2 = 0.19$ and $\Delta =0.19$. We observe that the oscillation pattern could be identified more clearly. As mentioned in the main text, we have further added a non-zero detuning $\Delta$, in order to offset the influence induced by $U_2$. Through the simple grid search optimization of the parameters $\Omega$ and $\Delta$, in plots (c), (f), among several instances with relatively high oscillation contrast, we display the one with parameters $\Omega =2$, $U_1 =12$, $U_2 = 0.19$ and $\Delta =0.38$, which is the 2D view of Fig. \ref{fig:experiment} in the main text. The linear light cone contour and periodic oscillations inside the light cone can be readily observed.

In the main text, we have numerically calculated the $X Z$-OTOC dynamics of the PXP model. However, $X Z$-OTOCs can not be directly reduced to the observable average form like Eq. \eqref{Eq:ob_ave_exp} in the main text, thus do not have a similar measurement scheme like the $Z Z$-OTOCs. Several other methods such as the ones based on randomized measurements \cite{Vermersch2019Probing} or classical shadow estimations \cite{Garcia2021Quantum} might be modified and adopted instead.

\bibliographystyle{apsrev4-1-title}
\bibliography{DengQAIGroup,QMBS}

\end{document}